\documentclass[12pt]{article}
\begin{document}
\vspace*{2cm} \noindent {\Large\bf Large Deviations on a Cayley
Tree I:\newline Rate Functions.} \vspace{\baselineskip}
\newline {\bf A.E.\ Patrick\footnote[1]{Laboratory of Theoretical
Physics, Joint Institute for Nuclear Research, Dubna 141980,
Russia\\ e-mail: patrick@theor.jinr.ru}}
\begin{list}{}{\setlength{\rightmargin }{0mm}
\setlength{\leftmargin }{2.5cm}} \item \rule{124mm}{0.3mm}
\linebreak {\footnotesize {\bf Abstract.} We study the spherical
model of a ferromagnet on a Cayley tree and show that in the case
of empty boundary conditions the ferromagnetic phase transition
takes place at the critical temperature
$T_c=\frac{6\sqrt{2}}{5}J$, where $J$ is the interaction strength.
For any temperature the equilibrium magnetization, $m_n$, tends to
zero in the thermodynamic limit, and the true order parameter is
the renormalized magnetization $r_n=n^{3/2}m_n$, where $n$ is the
number of generations in the Cayley tree. Below $T_c$, the
equilibrium values of the order parameter are given by
\[
\rho^*=\pm\frac{2\pi}{(\sqrt{2}-1)^2}\sqrt{1-\frac{T}{T_c}}.
\]
There is one more notable temperature, $T_{\rm p}$, in the model.
Below that temperature the influence of homogeneous boundary field
penetrates throughout the tree. We call $T_{\rm p}$ the
penetration temperature, and it is given by
\[
T_{\rm p}=\frac{J}{W_{\rm
Cayley}(3/2)}\left(1-\frac{1}{\sqrt{2}}\left(\frac{h}{2J}\right)^2
\right).
\]
The main new technical result of the paper is a complete set of
orthonormal eigenvectors for the discrete Laplace operator on a
Cayley tree.}
\newline \rule[1ex]{12.4cm}{0.3mm}
\linebreak {\bf \sc key words:} {\footnotesize Critical
temperature; order parameter; phase transition; spherical model}
\vspace{\baselineskip}
\end{list}

\section{Introduction.}

It is well known that some thermodynamic observables of the Ising
model on a Cayley tree (the IC model) are non-analytic functions
of the temperature. In the case of a tree with branching ratio
$k$, the expected values of microscopic variables (local
magnetization) are singular at $T_{\rm B}\equiv\beta_{\rm
B}^{-1}$: $\tanh(J \beta_{\rm B})=1/k$. However, the
susceptibility of total magnetization diverges for $T\leq T_{\rm
C}\equiv\beta_{\rm C}^{-1}$: $\tanh(J \beta_{\rm C})=1/\sqrt{k}$,
see \cite{m_1974} and \cite{mh_1975}. The very existence of two
ferromagnetic critical points is rather puzzling, because in
finite-dimensional systems diverging susceptibility is usually
accompanied by steeply rising spontaneous magnetization. The
present paper is an attempt to shed some light on the mystery of
this double critical-point phenomenon.

One of the distinctive features of Cayley trees is abnormally
large number of boundary sites. In fact, the boundary is a
macroscopic part of entire tree. Therefore, it does not come as a
surprise, that properties of Cayley-tree models are most
intriguing in the case of zero boundary field (the empty boundary
conditions), see, e.g., \cite{brz_1995, m_1974}. For example,
below the critical temperature the distribution of total
magnetization in the 2D Ising model or in the 3D spherical model
with empty boundary conditions concentrates around two points
$\pm m^*(T)$, where $m^*(T)$ is the spontaneous magnetization.
On the contrary, the distribution of magnetization in the IC
model always concentrates around zero. That is, the total
spontaneous magnetization in the IC model is equal to zero
on the entire interval $T<T_{\rm C}$ where the susceptibility
is infinite.

In order to resolve a paradoxical case or a seemingly
contradictory situation it is important to present all available
facts in a most clear and transparent way. To obtain such a
description of important thermodynamic properties of a model (with
or) without symmetry-breaking perturbations one can look at
large-deviation probabilities of thermodynamic observables
\[
\Pr[m_N\in[a,b]]\sim\exp\left[-N\min_{x\in[a,b]}R(x)\right].
\]
The typical behavior of rate functions $R(x)$ within many models
of statistical mechanics can be summarized by the following
standard scenario. If the temperature is sufficiently high, the
function $R(x)$ has a positive second derivative and a unique
minimum (in fact, zero) at the equilibrium value $m(T)$ of the
variable $m_N$ under consideration. However, when the temperature
drops to a critical value, $T_{\rm c}$, the second derivative at
the point of minimum vanishes, $R''(m(T_{\rm c}))=0$. If the
temperature is reduced even further, then the minimum of $R(x)$
either stretches into a flat horizontal segment or, in the case of
mean-field models, splits into several points of minima. In the
former case the low-temperature phases are called soft, in the
later case the phases are called rigid.

Another surprising (in view of the above standard scenario)
property of Cayley-tree models was reported in the
paper \cite{brssz_2001}, where it was shown that the second
derivative of the rate function $R''_{\rm IC}(x)$, describing
large-deviation probabilities of magnetization, does not vanish
neither at nor below the critical point. The second derivative of
a rate function at the minimum point determines the variance of
fluctuations of the variable under consideration around its
equilibrium value. When the second derivative tends to zero, the
variance of fluctuations tends to infinity signaling that the
thermodynamic system is approaching a critical point. Therefore,
non-vanishing second derivative of rate functions in Cayley-tree
models raises a question on the nature of phase transitions there.

The IC model has a close relative --- the Ising model on a Bethe
lattice (the IB model). A derivation of the exact formula for the
magnetization of IB model can be found, for instance, in Section 4
of the famous book by Baxter \cite{b_1982}. What was also derived
in \cite{b_1982} is the free energy of IB model in the ensemble
with fixed magnetization, whence the rate function of
magnetization, $R_{\rm IB}(x)$, can be extracted. Although the
status of thus derived expressions is not quite clear, the
obtained rate function $R_{\rm IB}(x)$ also exhibits a surprising
feature established rigorously in \cite{brssz_2001}
--- the second derivative of $R_{\rm IB}(x)$ is strictly positive
for all temperatures. Moreover, $R''_{\rm IB}(x)$ at the
equilibrium value $x=0$ becomes a linear function of $\beta J$,
$R''_{\rm IB}(0)=\frac{1}{9}\beta J$, for $\beta>\beta_{\rm B}$.

From a geometric point of view, a Cayley tree is a kind of a loose
bundle of 1D lattices. Can such a bundle exhibit a phase
transition and support true criticality associated with emerging
long-range correlations between macroscopically separated domains?
It is quite conceivable that the singularity in the IB model has
nothing to do with strong correlations. Instead, the critical
point might signal penetration inside the tree of an effective
field induced by boundary conditions imposed on a macroscopically
large part of the lattice and its accumulation in a mesoscopic
domain inside the tree. However, as far as some other models on
Cayley trees are concerned, an evidence of strong correlations was
established in \cite{m_1974}, where it was shown that the
susceptibility of the IC model diverges at $T=T_{\rm C}$, and in
\cite{bdp_1992}, where it was shown that the free boson gas on a
Cayley tree exhibits condensation in the ground state for
sufficiently low temperatures (although, the exact value of the
critical temperature was not reported in \cite{bdp_1992}).

Taking into account a host of intriguing features found in
Cayley-tree models, it seems worthwhile to investigate the
properties of the corresponding spherical model. First of all, it
is interesting to find out which of the features found in models
with discrete microscopic variables are also present in continuous
models. Second, owing to the gaussian distribution of the
microscopic random variables (above the critical temperature, in
any case), one might hope to obtain a more complete and explicit
description of thermodynamic properties of the spherical model,
than the results obtained for the IC model  so far. It turns out,
that properties of the spherical and the Ising models on a Cayley
tree are quite similar. In particular, the second derivative of
the rate function $R_{\rm Sph}(x)$ describing large-deviation
probabilities of the magnetization in the spherical model is also
positive at the unique minimum $x=0$ for any temperature.
Moreover, analogously to $R''_{\rm IB}(0)$, the second derivative
of $R_{\rm Sph}(x)$ at $x=0$ also simplifies to a linear function
of $\beta J$ for $\beta>\beta_c$:
\[
R''_{\rm Sph}(0)=\frac{(\sqrt{2}-1)^2}{\sqrt{2}}\beta J.
\]

Most likely, the behavior of rate functions in Cayley-tree models
follows the mean-field scenario if we look at large-deviation
probabilities of appropriate order parameters. In the case of the
spherical model, the behaviour of the rate function $R_{\rm Cayley}(x)$
describing large-deviation probabilities of the renormalized magnetization,
\[
r_{N,\frac{3}{2}}=(\log_2 N)^\frac{3}{2}m_{N}\equiv
\frac{(\log_2 N)^\frac{3}{2}}{N}\sum_{(j,k)\in T_n}x_{j,k},
\]
does follow the mean-field scenario. Namely, the unique minimum of
$R_{\rm Cayley}(x)$ at $x=0$
splits in two isolated points of minima $x=\pm \rho^*$, and the
rate function is no longer convex when the temperature falls
below $T_{\rm c}=\frac{6\sqrt{2}}{5}J$. Nevertheless, the rate function $R_{\rm
Cayley}(x)$ is also not devoid of unusual properties. As the
temperature drops to its critical value $T_{\rm c}$, the second
derivative $R''_{\rm Cayley}(0)$ does not vanish, but tends to the
positive value $\frac{5}{12}(\sqrt{2}-1)^2$.

The rest of the paper is organized as follows. In Section 2 we
define the spherical model on a Cayley tree and present several
technical results that are used in the later sections. The main
results of the paper are summarized in Section 3. In Section 4 we
derive the main asymptotics of the free energy and establish the
phase diagrams of the spherical model with three boundary conditions:
empty, homogeneous, and alternating (antiferromagnetic). Section 5 is devoted to an
investigation of large-deviation probabilities for
magnetization. The ground-state (zero-temperature) properties
of the spherical model are investigated in Section 6.
In Section 7 we look at the large-deviation probabilities of the
renormalized magnetization $r_{N,\frac{3}{2}}$
--- the true order parameter of the spherical model on a Cayley tree.
The results of the paper are discussed in Section 8.

\section{The model and useful facts.}

Consider a binary Cayley tree $T_n$ --- a tree with branching ratio two containing
$n$ generations. Each node of the tree is labelled by a pair of integers
$(k,l)$, where the first integer indicates the tree generation,
$k=1,2,\ldots,n$, and the second integer numbers nodes within the
$k$-th generation, $l=1,2,\ldots,2^{k-1}$, see Fig.\ 1. There are exactly
$N=2^n-1$ nodes in a tree with $n$ generations. The spherical model
on a Cayley tree describes a collection of random variables
$\{x_{j,k}:(j,k)\in T_n\}$ attached to the nodes of the tree $T_n$.

\begin{figure}[t]
\setlength{\unitlength}{1mm}
\begin{picture}(150,70)(3,-10)

\put(42,53){\bf 1${}^{\rm st}$ generation} \put(80,54){ $x_{1,1}$}

\multiput(77.2,54)(1.46,-0.6){26}{\makebox(0,0){\circle*{0.2}}}
\multiput(77.2,54)(-1.46,-0.6){26}{\makebox(0,0){\circle*{0.2}}}

\multiput(40.7,39)(1,-0.9){19}{\makebox(0,0){\circle*{0.2}}}
\multiput(40.7,39)(-1,-0.9){19}{\makebox(0,0){\circle*{0.2}}}
\put(44.5,38){$x_{2,1}$} \put(117.5,38){$x_{2,2}$}

\multiput(113.7,39)(1,-0.9){19}{\makebox(0,0){\circle*{0.2}}}
\multiput(113.7,39)(-1,-0.9){19}{\makebox(0,0){\circle*{0.2}}}

\put(5,38){\bf 2${}^{\rm nd}$ generation}

\multiput(21.7,21.9)(1,-1.1){10}{\makebox(0,0){\circle*{0.2}}}
\multiput(21.7,21.9)(-1,-1.1){10}{\makebox(0,0){\circle*{0.2}}}
\multiput(59.7,21.9)(1,-1.1){10}{\makebox(0,0){\circle*{0.2}}}
\multiput(59.7,21.9)(-1,-1.1){10}{\makebox(0,0){\circle*{0.2}}}
\multiput(94.7,21.9)(1,-1.1){10}{\makebox(0,0){\circle*{0.2}}}
\multiput(94.7,21.9)(-1,-1.1){10}{\makebox(0,0){\circle*{0.2}}}
\multiput(132.7,21.9)(1,-1.1){10}{\makebox(0,0){\circle*{0.2}}}
\multiput(132.7,21.9)(-1,-1.1){10}{\makebox(0,0){\circle*{0.2}}}

\put(24.5,20.9){$x_{3,1}$} \put(62.5,20.9){$x_{3,2}$}
\put(97.5,20.9){$x_{3,3}$} \put(135.5,20.9){$x_{3,4}$}

\multiput(12.7,12)(0.6,-1.3){10}{\makebox(0,0){\circle*{0.2}}}
\multiput(12.7,12)(-0.6,-1.3){10}{\makebox(0,0){\circle*{0.2}}}

\multiput(30.7,12)(0.6,-1.3){10}{\makebox(0,0){\circle*{0.2}}}
\multiput(30.7,12)(-0.6,-1.3){10}{\makebox(0,0){\circle*{0.2}}}

\multiput(50.7,12)(0.6,-1.3){10}{\makebox(0,0){\circle*{0.2}}}
\multiput(50.7,12)(-0.6,-1.3){10}{\makebox(0,0){\circle*{0.2}}}

\multiput(68.7,12)(0.6,-1.3){10}{\makebox(0,0){\circle*{0.2}}}
\multiput(68.7,12)(-0.6,-1.3){10}{\makebox(0,0){\circle*{0.2}}}

\multiput(85.7,12)(0.6,-1.3){10}{\makebox(0,0){\circle*{0.2}}}
\multiput(85.7,12)(-0.6,-1.3){10}{\makebox(0,0){\circle*{0.2}}}

\multiput(103.7,12)(0.6,-1.3){10}{\makebox(0,0){\circle*{0.2}}}
\multiput(103.7,12)(-0.6,-1.3){10}{\makebox(0,0){\circle*{0.2}}}

\multiput(123.7,12)(0.6,-1.3){10}{\makebox(0,0){\circle*{0.2}}}
\multiput(123.7,12)(-0.6,-1.3){10}{\makebox(0,0){\circle*{0.2}}}

\multiput(141.7,12)(0.6,-1.3){10}{\makebox(0,0){\circle*{0.2}}}
\multiput(141.7,12)(-0.6,-1.3){10}{\makebox(0,0){\circle*{0.2}}}

\put(15.5,11){$x_{4,1}$} \put(33.5,11){$x_{4,2}$}
\put(53.5,11){$x_{4,3}$} \put(71.5,11){$x_{4,4}$}
\put(88.5,11){$x_{4,5}$} \put(106.5,11){$x_{4,6}$}
\put(126.5,11){$x_{4,7}$} \put(144.5,11){$x_{4,8}$}

\put(146.8,0.3){{\circle*{1}}} \put(136.1,0.3){{\circle*{1}}}
\put(128.8,0.3){{\circle*{1}}} \put(118.1,0.3){{\circle*{1}}}
\put(108.8,0.3){{\circle*{1}}} \put(98.1,0.3){{\circle*{1}}}
\put(90.8,0.3){{\circle*{1}}} \put(80.1,0.3){{\circle*{1}}}

\put(77,54){{\circle*{1}}}

\put(40.5,39){{\circle*{1}}} \put(113.5,39){{\circle*{1}}}

\put(21.5,21.9){{\circle*{1}}} \put(59.5,21.9){{\circle*{1}}}
\put(94.5,21.9){{\circle*{1}}} \put(132.5,21.9){{\circle*{1}}}

\multiput(123.5,12)(18,0){2}{{\circle*{1}}}
\multiput(85.5,12)(18,0){2}{{\circle*{1}}}
\multiput(50.5,12)(18,0){2}{{\circle*{1}}}
\multiput(12.5,12)(18,0){2}{{\circle*{1}}}
\multiput(7,0.3)(18,0){2}{{\circle*{1}}}
\multiput(17.8,0.3)(18,0){2}{{\circle*{1}}}
\multiput(45.2,0.3)(18,0){2}{{\circle*{1}}}
\multiput(56,0.3)(18,0){2}{{\circle*{1}}}

\put(5,-4){$x_{5,1}$} \put(15,-4){$x_{5,2}$}
\put(23,-4){$x_{5,3}$} \put(33,-4){$x_{5,4}$}

\put(43,-4){$x_{5,5}$} \put(53.5,-4){$x_{5,6}$}
\put(61.5,-4){$x_{5,7}$} \put(70,-4){$x_{5,8}$}

\put(78.5,-4){$x_{5,9}$} \put(87,-4){$x_{5,10}$}
\put(96.5,-4){$x_{5,11}$} \put(106.5,-4){$x_{5,12}$}

\put(116.5,-4){$x_{5,13}$} \put(126.1,-4){$x_{5,14}$}
\put(135,-4){$x_{5,15}$} \put(145,-4){$x_{5,16}$}

\end{picture}
\caption{A binary Cayley tree $T_5$ --- a tree with branching ratio
two containing five generations. A random variable $x_{j,k}$ is attached
to each node $(j,k)$ of the tree. Nearest-neighbour nodes are connected
by dotted lines. }
\end{figure}
\vspace{\abovedisplayskip}

\underline{\bf The Hamiltonian} \vspace{\abovedisplayskip}

\noindent The interaction of the variables $x_{j,k}$
is described by the Hamiltonian
\begin{equation}
H_n=-J\sum_{(j,k),(l,m)\in T_n}M_{(j,k),(l,m)}x_{j,k} x_{l,m}-
\sum_{k=1}^{2^{n-1}}h_k\, x_{n,k},
\label{ham}
\end{equation} where $J>0$,
\begin{equation}
M_{(j,k),(l,m)}=\left\{
\begin{array}{cl}\frac{1}{2},&\mbox{ if }l=j+1,m\in\{2k-1,2k\},\vspace{1mm}\\
\frac{1}{2},&\mbox{ if }l=j-1,k\in\{2m-1,2m\},\vspace{1mm}\\
0,&\mbox{ otherwise,}
\end{array}
\right.
\label{ctm}
\end{equation}
are elements of the symmetric (nearest neighbour) Cayley-tree
matrix $\widehat M_N$, and $\{h_k:k=1,2,\ldots, 2^{n-1}\}$ is a
boundary field. Note that according to the Hamiltonian $H_n$ only
variables $x_{j,k}$ located at nearest-neighbour nodes of the
Cayley tree interact with each other directly. In this paper we
consider three types of boundary conditions: $h_k=0$ (empty
b.c.),\ $h_k=h$ (homogeneous b.c.),\ and $h_k=(-1)^k h, \mbox{ for
} k=1,2,\ldots, 2^{n-1}$ (alternating b.c.).

The $N$ eigenvalues, $\{\lambda_{j,k}:(j,k)\in T_n\}$, of the
matrix $\widehat M_N$ and the corresponding eigenvectors
$\{\mbox{\boldmath $u$}^{(j,k)}:(j,k)\in T_n\}$ are found in
Appendix A. The spectrum of the matrix $\widehat M_N$ (the set of
different eigenvalues), $\{\tau_{j,k}\}_{k=1,}^j{}_{j=1}^n$,
contains exactly $n(n+1)/2$ real numbers. It is convenient to
arrange the values $\tau_{j,k}$ and their multiplicities $m_{j,k}$
in triangular arrays
\begin{equation}
\begin{array}{lcl}
\tau_{1,1}(=0) && m_{1,1}=2^{n-2}\\
\tau_{2,1},\tau_{2,2} && m_{2,1}=m_{2,2}=2^{n-3}\\
\tau_{3,1},\tau_{3,2},\tau_{3,3}&& m_{3,1}=m_{3,2}=m_{3,3}=2^{n-4}\\
\qquad\vdots&\qquad&\qquad\vdots\\
\tau_{n-1,1},\tau_{n-1,2},\ldots,\tau_{n-1,n-1}&&
m_{n-1,1}=
m_{n-1,2}=\ldots=m_{n-1,n-1}=1\\
\tau_{n,1},\tau_{n,2},\tau_{n,3},\ldots,\tau_{n,n-1},\tau_{n,n}&&
m_{n,1}= m_{n,2}=m_{n,3}=\ldots=m_{n,n}=1
\end{array}
\label{evals}
\end{equation}
In fact, the $k$-th line of the above triangular array contains
the eigenvalues
\[
\Lambda_{k;l}=\sqrt{2}\cos\frac{\pi l}{k+1},\quad l=1,2,\ldots,k
\]
of the $k\times k$ tri-diagonal matrix
\[
\widehat L_k=\left(
\begin{array}{ccccccc}0&1&&&&&\\
\frac{1}{2}&0&1&&&\mbox{\LARGE 0}&\\
&\frac{1}{2}&0&\ddots&&&\\
&&\ddots&\ddots&\ddots&&\\
&&&\ddots&0&1&\\
&\mbox{\LARGE 0}&&&\frac{1}{2}&0&1\\
&&&&&\frac{1}{2}&0
\end{array}
\right).
\]

The normalized eigenvectors $\mbox{\boldmath $v$}^{(n,l)}$ corresponding
to the last line of eigenvalues $\tau_{n,l}$ in Eq.\ (\ref{evals}) are given by
\begin{equation}
\left\{v_{k,m}^{(n,l)}\right\}_{(k,m)\in
T_n}=\left\{\frac{2^{1-k/2}}{\sqrt{n+1}}\sin\frac{\pi
lk}{n+1}\right\}_{(l,m)\in T_n}.
\label{sev}
\end{equation}
Note that the components $v_{k,m}^{(n,l)}$ do not depend on $m$,
that is, they are identical within each generation of the tree.
Below, we refer to these vectors $\mbox{\boldmath $v$}^{(n,l)}$ as
special eigenvectors.\vspace{\abovedisplayskip}

\underline{\bf The Gibbs distribution} \vspace{\abovedisplayskip}

The joint distribution of the random variables $\{x_{j,k}:(j,k)\in
T_n\}$ is specified by the usual Gibbs density
\[
p(\{x_{j,k}:(j,k)\in T_n\})=\frac{e^{-\beta H_n}}{\Theta_n},
\]
with respect to the spherical {\em ``a priori''} measure
\[
\mu_n(dx)=\delta\left(\sum_{(j,k)\in
T_n}x_{j,k}^2-N\right)\prod_{(j,k)\in T_n}dx_{j,k}.
\]
The normalization factor (partition function) $\Theta_n$ is given
by
\begin{equation}
\Theta_n=\int_{-\infty}^\infty\ldots\int_{-\infty}^\infty
e^{-\beta H_n}\mu_n(dx). \label{pf}
\end{equation}

\underline{\bf Useful sums and asymptotic expansions} \vspace{\abovedisplayskip}

\noindent Using the explicit expressions for the spectrum of the
Cayley-tree matrix and the corresponding multiplicities, see Eq.\
(\ref{evals}), we obtain
\begin{eqnarray}
L_n(z)&\equiv&\frac{1}{N}\sum_{(j,k)\in T_n}\ln(z-\lambda_{j,k}) \label{lnz}\\
&=&
\frac{1}{N}\sum_{j=1}^{n-1}\sum_{k=1}^{j}2^{n-1-j}\ln\left(z-\sqrt{2}\cos\frac{\pi
k}{j+1}\right)+\frac{1}{N}\sum_{k=1}^{n}\ln\left(z-\sqrt{2}\cos\frac{\pi
k}{n+1}\right).\nonumber
\end{eqnarray}

The identity
\begin{equation}
\sum_{k=1}^{j}\ln\left(z-\sqrt{2}\cos\frac{\pi
k}{j+1}\right)=(j+1)\ln
\frac{x_+(z)}{\sqrt{2}}+\ln\frac{1-x_-^{2(j+1)}(z)}{\sqrt{z^2-2}},
\label{slog}
\end{equation}
where
\[
x_{\pm}(z)=\frac{z\pm\sqrt{z^2-2}}{\sqrt{2}},
\]
see, e.g., \cite{p94}, allows one to get rid of the summations
over $k$. As a consequence, for $z>\sqrt{2}$ we obtain the following large-$n$
asymptotic expansion
\begin{equation}
\frac{1}{N}{\sum_{(j,k)\in
T_n}}\ln(z-\lambda_{j,k})=L(z)+O\left(\frac{\log_2 N}{N}\right),
\label{sln}
\end{equation}
where
\begin{equation}
L(z)
=\frac{3}{2}\ln\frac{x_+(z)}{\sqrt{2}}-\frac{1}{4}\ln\left(z^2-2\right)+
\sum_{j=2}^\infty
2^{-j}\ln\left[1-x_-^{2j}(z)\right].\label{lz}
\end{equation}

The definition of $L(z)$ requires clarifications when
$z\leq\sqrt{2}$. In this case instead of Eqs.\ (\ref{sln}) and
(\ref{lz}) one can use the following ``accelerated" asymptotic
expansion
\[
\frac{1}{N}{{\sum\,}'_{(j,k)\in
T_n}}\ln\left(\tau_{n-1,1}+\frac{\zeta}{n^3}-\lambda_{j,k}\right)=A_{n-1}+\frac{\zeta}{n^3} B_{n-1}+
O(n^{-6}),
\]
where the prime indicates that the sum over $(j,k)$ does not
include the term corresponding to the maximal eigenvalue
$\tau_{n,1}$ and
\begin{eqnarray}
A_n&=&-\frac{1}{2}\ln 2+\sum_{j=2}^n 2^{-j}\ln\frac{\sin\frac{\pi
j}{n+1}}{\sin\frac{\pi}{n+1}}, \label{an}\\
B_n&=&\frac{1}{\sin\frac{\pi}{n+1}}\sum_{j=2}^n
2^{-j}\left(\cot\frac{\pi }{n+1}-j\cot\frac{\pi
j}{n+1}\right)=\frac{5}{3}+O(n^{-2}).\nonumber
\end{eqnarray}

Finally, if $z=\tau_{n,1}+\frac{\zeta}{N}-\lambda_{j,k}$, then the large-$n$ asymptotic expansion
of $L_n(z)$ is given by
\begin{equation}
\frac{1}{N}{\sum_{(j,k)\in
T_n}}\ln\left(\tau_{n,1}+\frac{\zeta}{N}-\lambda_{j,k}\right)=
C_n+\frac{1}{N}\ln\frac{\zeta}{N}+B_n
\frac{\zeta}{N}+O\left(N^{-2}\right), \label{sln1}
\end{equation}
where
\[
C_n=A_n+\frac{1}{N}\ln\frac{n+1}{\sin^2\frac{\pi}{n+1}}.
\]

The following sum (it can be calculated using, for instance, the
``contour summation" technique, see \cite{p94}) will also prove
useful
\begin{equation}
\frac{1}{n+1}\sum_{k=1}^n \frac{\sin\frac{\pi n
k}{n+1}\sin\frac{\pi j k}{n+1}}{z-\sqrt{2}\cos\frac{\pi k}{n+1}}=
\frac{1}{\sqrt{2}}\frac{x_+^{j}(z)-x_-^{j}(z)}{x_+^{n+1}(z)-x_-^{n+1}(z)}.
\label{us1}
\end{equation}

\section{The main results.}

The main theme of the present paper are thermodynamic properties
of the spherical model on a Cayley tree with branching ratio 2.
The important results can be stated as follows.
\begin{enumerate}
\item In the case of empty boundary conditions the spherical model
transits into a ferromagnetic state at the critical temperature
$T_c=\frac{6\sqrt{2}}{5}J$.

\item Since nearly half of Cayley-tree sites belong to the
boundary, the value and the very existence of the critical
temperature depends on the type of boundary conditions imposed. In
the spherical model with the alternating boundary field
$h_{n,k}=(-1)^k h$, the critical temperature exists only if
$|h|<4J$, and it is given by
\[
T_c(h)=\left[1-\left(\frac{h}{4J}\right)^2\right]\frac{6\sqrt{2}}{5}J.
\]

\item For any temperature $T$, the rate function $R_{\rm Sph}(x)$
describing large-deviation probabilities of the magnetization
\[
\Pr\left[\frac{1}{N}\sum_{(j,k)\in T_n}x_{j,k}\in[a,b]\right]
\sim\exp\left[-N\min_{x\in[a,b]}R_{\rm Sph}(x)\right]
\]
is a strictly convex analytic function vanishing only at $x=0$.
The second derivative $R_{\rm Sph}''(0)$ at the minimum point is
always positive and
\[
R_{\rm Sph}''(0)=\frac{(\sqrt{2}-1)^2}{\sqrt{2}}\beta J,
\]
for $T<T_c$.

\item The true order parameter of the spherical model is the
renormalized magnetization
\[
r_{N,\frac{3}{2}}=(\log_2 N)^\frac{3}{2}m_{N}\equiv \frac{(\log_2
N)^\frac{3}{2}}{N}\sum_{(j,k)\in T_n}x_{j,k}.
\]
Below the critical temperature $T_c$ the equilibrium values of
$r_{N,\frac{3}{2}}$ --- zeroes of the corresponding  rate function
$R_{\rm Cayley}(\rho)$ --- are given by $\rho=\pm\rho^*$, where
\[
\rho^*\equiv
\frac{2\pi}{(\sqrt{2}-1)^2}\sqrt{1-\frac{\beta_c}{\beta}}.
\]
The rate function $R_{\rm Cayley}(\rho)$ is not convex, therefore,
the low-temperature phases of the spherical model on a Cayley-tree
are of mean-field type and rigid, in the terminology
of the paper \cite{brssz_2001}.
%

\item Homogeneous boundary conditions generate an effective field
that penetrates toward the tree root if $T<T_p$. The penetration
temperature is an analogue of the critical temperature of the
Ising model on a Bethe lattice and it is given by
\[
T_{\rm p}=\frac{J}{W_{\rm
Cayley}(3/2)}\left(1-\frac{1}{\sqrt{2}}\left(\frac{h}{2J}\right)^2
\right),
\]
where $W_{\rm Cayley}(z)$ is define in Eq.\ (\ref{wc}).
\end{enumerate}

\section{The phase diagram.}
The partition function $\Theta_n$ of the spherical model on a
Cayley tree can be calculated using the technique described in the
famous paper by Berlin and Kac \cite{bk_1952}. Here we outline
only the main steps.

Introduction of new integration variables $\{y_{l,m}:(l,m)\in
T_n\}$ in Eq.\ (\ref{pf}) via
\[
x_{j,k}=\sum_{(l,m)\in T_n}u_{j,k}^{(l,m)}y_{l,m},\quad (j,k)\in
T_n,
\]
where $\{\mbox{\boldmath $u$}^{(l,m)}:(l,m)\in T_n\}$ are
orthonormal eigenvectors of the Caley-tree matrix $\widehat M_N$,
diagonalises the Hamiltonian (\ref{ham}). Therefore, we obtain the
following formula for the partition function
\[
\Theta_n=\int_{-\infty}^\infty\ldots\int_{-\infty}^\infty
\exp\left[\beta J\sum_{(l,m)\in T_n}\lambda_{l,m}
y_{l,m}^2+\beta\sum_{(l,m)\in
T_n}\phi_{l,m}y_{l,m}\right]\mu_n(dy),
\]
where $\{\lambda_{l,m}:(l,m)\in T_n\}$ are the eigenvalues of the
matrix $\widehat M_N$, see Eq.\ (\ref{evals}), and
\begin{equation}
\phi_{l,m}=\sum_{k=1}^{2^{n-1}}u_{n,k}^{(l,m)}h_k
\label{hphi}
\end{equation}
are ``scalar products" of the boundary field and the eigenvectors of
the matrix $\widehat M_N$.

The integral representation
for the delta function in the {\em ``a priori''} measure,
\[
\delta\left(\sum_{(j,k)\in T_n}y_{j,k}^2-N\right)=\frac{1}{2\pi
i}\int_{-i\infty}^{+i\infty}\!ds\,
\exp\left[s\left(N-\sum_{(j,k)\in T_n}y_{j,k}^2\right)\right],
\]
allows one to perform
integration over the new variables $\{y_{j,k}:(j,k)\in T_n\}$.
However, one can switch the order of integration over the
variables $y_{j,k}$ and $s$ only after a shift of the integration
contour for $s$ to the right. The shift must assure that the real
part of the quadratic form involving the variables $y_{j,k}$ is
negatively defined. Switching the integration order, integrating
over $\{y_{j,k}:(j,k)\in T_n\}$, and introducing a new integration
variable $z$ via $s=\beta J z$ we obtain
\begin{equation}
\Theta_n=\frac{\beta J}{2\pi i}\left(\frac{\pi}{\beta
J}\right)^{N/2}\int_{-i\infty+c}^{+i\infty+c}\!dz\,
\exp\left[N\beta\Phi_n(z)\right], \label{pf1}
\end{equation}
where
\[
\Phi_n(z)=J z-\frac{1}{2\beta N}\sum_{(j,k)\in
T_n}\ln(z-\lambda_{j,k})+\frac{1}{4 J N}\sum_{(j,k)\in
T_n}\frac{\phi_{j,k}^2}{z-\lambda_{j,k}},
\]
and $c>\sqrt{2}$ is the shift of integration contour
mentioned above.

The large-$n$ asymptotics of the integral (\ref{pf1}) can be found
using the saddle-point method. In the case of empty boundary
conditions all scalar products $\phi_{j,k}$ are equal to zero, and
the saddle point of the integrand is a solution of the equation
\begin{equation}
\Phi_n'(z)=0\quad \Leftrightarrow \quad J -\frac{1}{2\beta
N}\sum_{(j,k)\in T_n}\frac{1}{z-\lambda_{j,k}}=0. \label{sp1}
\end{equation}
Making use of the explicit expressions for the spectrum
$\tau_{j,k}$ of the matrix $\widehat M_N$ and the corresponding
multiplicities $m_{j,k}$, see Eq.\ (\ref{evals}), we obtain
\[
J -\frac{1}{2\beta
N}\left[\sum_{j=1}^{n-1}2^{n-1-j}\sum_{k=1}^j\frac{1}{z-\sqrt{2}\cos\frac{\pi
k}{j+1}}+\sum_{k=1}^n\frac{1}{z-\sqrt{2}\cos\frac{\pi
k}{n+1}}\right]=0.
\]

Differentiation of Eq.\ (\ref{slog}) over $z$ yields
\[
\sum_{k=1}^j\frac{1}{z-\sqrt{2}\cos\frac{\pi
k}{j+1}}=\frac{j+1}{\sqrt{z^2-2}}\left(1+\frac{2}{x_+^{2(j+1)}(z)-1}\right)-\frac{z}{z^2-2}.
\]
Therefore, assuming $z>\sqrt{2}$ and passing to the limit
$n\to\infty$ we obtain the following equation for the saddle-point
$z^*$:
\begin{equation}
\Phi'(z)\equiv J-\frac{1}{2\beta}W_{\rm Cayley}(z) =0,
\label{lspe}
\end{equation}
where
\begin{equation}
W_{\rm Cayley}(z)=\sum_{j=2}^{\infty}
2^{-j}\left[\frac{j}{\sqrt{z^2-2}}\left(1+\frac{2}{x_+^{2j}(z)-1}\right)-\frac{z}{z^2-2}
\right]
\label{wc}
\end{equation}
is the Cayley-tree analogue of the Watson function from the
original paper on the spherical model by Berlin and Kac, see
\cite{bk_1952}.

The function $\Phi'(z)$ increases monotonically with $z$ on
$(\sqrt{2},\infty)$, and the location of its zeroes depends on the
value of the inverse temperature $\beta$. Since
\[
\Phi'(\sqrt{2})=\lim_{z\downarrow
\sqrt{2}}\Phi'(z)=J-\frac{5}{6\sqrt{2}\beta},
\]
there exists a critical value
\[
\beta_c =\frac{5}{6\sqrt{2}J}
\]
of the inverse temperature $\beta$. If $\beta\in(0,\beta_c)$
(high-temperatures), then the function $\Phi'(z)$ has exactly one
zero on the interval $(\sqrt{2},\infty)$ at a point
$z^*>\sqrt{2}$. While if $\beta > \beta_c$ (low-temperatures),
then the function $\Phi'(z)$ is strictly positive for $z>
\sqrt{2}$.


The alternating boundary field $h_k=(-1)^k h$ is orthogonal to all
eigenvectors of the Cayley-tree matrix $\widehat M_N$ except for
the $2^{n-2}$ eigenvectors corresponding to the eigenvalue
$\tau_{1,1}=0$. A short calculation shows, that non-zero scalar
products (\ref{hphi}) are given by $\phi_{l,m}=\sqrt{2}h$.
Therefore, the saddle point of the integrand in Eq.\ (\ref{pf1})
is a solution of the equation
\begin{equation}
\Phi_n'(z)=0\quad \Leftrightarrow \quad J -\frac{1}{2\beta
N}\sum_{(j,k)\in
T_n}\frac{1}{z-\lambda_{j,k}}-\frac{1}{4JN}\frac{2h^2
2^{n-2}}{z^2}=0. \label{sp2}
\end{equation}
Assuming $z>\sqrt{2}$ and passing to the limit
$n\to\infty$ we obtain the following equation for the saddle-point
$z^*$:
\begin{equation}
\Phi'(z)\equiv J-\frac{1}{2\beta}W_{\rm Cayley}(z)-\left(\frac{h}{4J}\right)^2\frac{2}{z^2}=0.
\label{laspe}
\end{equation}
Thus, in the case of alternating boundary conditions the critical
temperature exists only if $|h|<4J$, and it is given by
\[
T_c^{\rm alt}(h)=
\left[1-\left(\frac{h}{4J}\right)^2\right]\frac{6\sqrt{2}J}{5}.
\]

If the boundary field is homogeneous, $h_k=h$, then its ``scalar
products" with the special eigenvectors $\mbox{\boldmath
$v$}^{(n,m)}$, see Eq.\ (\ref{hphi}), are given by
\[
\phi_{n,m}=\frac{2^{n/2}h}{\sqrt{n+1}}\sin\frac{\pi m
n}{n+1},\quad m=1,2,\ldots,n.
\]
All other eigenvectors of the Cayley-tree matrix are orthogonal to
a homogeneous boundary field. Therefore, Eq.\ (\ref{us1}) yields the
following formula for the function $\Phi_n(z)$ in the integral
(\ref{pf1})
\begin{equation}
\Phi_n(z)=Jz -\frac{1}{2\beta N}\sum_{(j,k)\in
T_n}\ln(z-\lambda_{j,k})+\frac{2^{n}h^2}{4JN\sqrt{2}}
\frac{x_+^{2n+1}(z)-x_+(z)}{x_+^{2n+2}(z)-1}. \label{sp3}
\end{equation}
Assuming $z>\sqrt{2}$, differentiating over $z$ and passing to the limit
$n\to\infty$ we obtain the following saddle-point equation
\begin{equation}
J -\frac{1}{2\beta}W_{\rm Cayley}(z)
-\frac{h^2}{8J}\frac{x_-(z)}{\sqrt{z^2-2}}=0. \label{spl3}
\end{equation}
As could have been expected, any homogeneous boundary field $h\neq
0$ keeps the saddle point $z^*$ away from the
singularity at $\sqrt{2}$ and, hence, destroys the phase
transition.

Application of the saddle-point method to the integral (\ref{pf1})
is straightforward when the saddle point $z^*$ is greater than
$\sqrt{2}$, see \cite{bk_1952}. In the case of alternating boundary
conditions, taking into account the estimate
(\ref{sln}) we obtain the following expression for the main
asymptotics of the partition function 
\begin{equation}
\Theta_n=\left(\frac{\pi}{\beta J}\right)^{N/2}
\exp\left[N\left(\beta Jz^*-\frac{1}{2}L(z^*)+\frac{\beta
h^2}{8Jz^*}\right)+O(n)\right], \label{maspfht}
\end{equation}
where $L(z)$ is given by Eq.\ (\ref{lz}). Analogously,
in the case of homogeneous boundary conditions we obtain
\begin{equation}
\Theta_n=\left(\frac{\pi}{\beta J}\right)^{N/2}
\exp\left[N\left(\beta Jz^*-\frac{1}{2}L(z^*)+\frac{\beta
h^2}{4J}\frac{1}{z^*+\sqrt{(z^*)^2-2}}\right)+O(n)\right], \label{maspfhbc}
\end{equation}
where $z^*$ is the solution of Eq.\ (\ref{spl3}) from the interval
$(\sqrt{2},\infty)$.

If $\beta\geq\beta_c^{\rm alt}(h)$, the function $\Phi_n(z)$ still
attains its minimum value on the interval $(\tau_{n,1},\infty)$ at
a point $z_n^*>\tau_{n,1}$, where $\tau_{n,1}\equiv
\sqrt{2}\cos\frac{\pi}{n+1}$ is the largest eigenvalue of the
Cayley-tree matrix $\widehat M_N$. However, now the sequence of
saddle points $z_n^*$ approaches the branch point of the integrand
at $z=\tau_{n,1}$, and application of the saddle-point method
becomes a bit more tricky. In fact, in the cases of empty and
alternating boundary conditions with $|h|<4J$, nothing prevents
the saddle point in Eq.\ (\ref{pf1}) from sliding towards the
branch point, and to find the main asymptotics of the partition
function $\Theta_n$ we have to introduce a new integration
variable $\zeta$ via $z=\tau_{n,1}+\zeta/N$. This change of
variables effectively eliminates the large parameter, $N$, in the
$\zeta$-dependent part of the integrand, and we are left with a
finite integral. Thus, the main asymptotics of the partition
function is given by
\begin{equation}
\Theta_n=\left(\frac{\pi}{\beta J}\right)^{N/2} \exp\left[N
\left(\beta J\tau_{n,1}-\frac{1}{2}A_n+\frac{\beta
h^2}{8J\tau_{n,1}}\right)+O(n)\right], \label{maspflt}
\end{equation}
where $A_n$ is given by Eq.\ (\ref{an}).

\section{Large Deviations of Magnetization.}

The main objective of this section is investigation of
large-deviation probabilities
\[
\Pr\left(\frac{1}{N}\sum_{(j,k)\in T_n}x_{j,k}\in[a;b]\right)
\]
for the magnetization --- the arithmetic average of microscopic
random variables $x_{j,k}$, $(j,k)\in T_n$. In particular, we
would like to find the main asymptotics of the distribution
densities
\begin{equation}
f_{n}(y)=\frac{1}{\Theta_n}\int^{+\infty}_{-\infty}\ldots\int^{+\infty}_{-\infty}
\delta\left(\frac{1}{N}\sum_{(j,k)\in
T_n}x_{j,k}-y\right)\exp[-\beta H_n]\mu_n(dx), \label{dd}
\end{equation}
as $N\to\infty$.

One can calculate the density $f_{n}(y)$ using the technique from
the previous section. This time, however, we have to deal with one
more delta function. Using the integral representation
\[
\delta(a)=\frac{1}{2\pi}\int_{-\infty}^{+\infty}\!e^{iaw}\,dw
\]
for the new delta function, and performing the integration over
the variables $x_{j,k}$ one obtains
\begin{eqnarray}
f_{n}(y)&=&\frac{N}{4\pi^2i\Theta_n}\int\limits_{-i\infty+c}
^{i\infty+c}\!ds\, e^{N s}\int\limits_{-\infty}^{\infty}
\!dw\,e^{i N y w}\times\nonumber\\
&\times&\prod_{(j,k)\in T_n}\sqrt{\frac{\pi}{s-\beta
J\lambda_{j,k}}} \exp\left[-\frac{\gamma_{j,k}^2 w^2}{4(s-\beta
J\lambda_{j,k})}\right], \label{tm1}
\end{eqnarray}
where
\[
\gamma_{j,k}\equiv\sum_{(l,m)\in T_n}u_{l,m}^{(j,k)}.
\]
Integrating over the variable $u$ and introducing a new
integration variable $z$ via $s=\beta Jz$ one arrives at
\begin{equation}
f_{n}(y)=\frac{\beta JN}{2\pi i\Theta_n}\left(\frac{\pi}{\beta J}
\right)^{(N-1)/2}\int\limits_{-i\infty+z_0}^{i\infty+z_0}\!\frac{dz}
{\sqrt{\Sigma_n(z)}}\exp\left[N\beta J\Phi_n(z,y)\right],
\label{tm2}
\end{equation}
where we have introduced the notations
\begin{equation}
\Sigma_n(z)=\sum_{(j,k)\in
T_n}\frac{\gamma_{j,k}^2}{z-\lambda_{j,k}}, \label{sgm}
\end{equation}
\begin{equation}
\Phi_n(z,y)=z-\frac{1}{2\beta JN}\sum_{(j,k)\in
T_n}\ln(z-\lambda_{j,k})- N\frac{y^2}{\Sigma_n(z)}. \label{fzm}
\end{equation}

If $z>\sqrt{2}$, then using Eq.\ (\ref{sss}) from Appendix B we
obtain the following limit
\begin{equation}
\Phi(z,y)\equiv\lim_{n\to\infty}\Phi_n(z,y)=z-\frac{1}{2\beta
J}L(z)-y^2\sigma(z), \label{lfzm}
\end{equation}
where
\[
\sigma(z)=\frac{2\left(z-\frac{3}{2}\right)^2}{3z-\sqrt{z^2-2}-4}.
\]
The Taylor expansion for $\sigma(z)$ at the point $z=\sqrt{2}$ is
given by
\[
\sigma(z)=\frac{3}{2\sqrt{2}}-1+2^{-5/4}\sqrt{z-\sqrt{2}}+\frac{3(z-\sqrt{2})}{4}+O\left[
\left(z-\sqrt{2}\right)^{3/2}\right].
\]
The term $\sqrt{z-\sqrt{2}}$ does not allow the saddle-points
$z_n(y)$ of $\Phi_n(z,y)$ to approach the singularity of the integrand at $z=\sqrt{2}$.
Therefore, provided $y\neq 0$, application of the saddle-point method to the integral (\ref{tm2})
is straightforward. For the main asymptotics of the distribution density we obtain
\[
f_{N}(y)=e^{-NR_{\rm Sph}(y)+O(\log_2 N)},
\]
where
\begin{equation}
R_{\rm Sph}(y)=\beta J(z^*-z^*(y))-
\textstyle{\frac{1}{2}}L(z^*)+\textstyle{\frac{1}{2}}L(z^*(y))
+\beta J y^2\sigma(z^*(y)), \label{lde1}
\end{equation}
is the rate function, $z^*(y)$ is the minimum point of the function
$\Phi(z,y)$ on the interval $[\sqrt{2};\infty)$, and $z^*=z^*(0)$.

The rate function $R_{\rm Sph}(y)$ has the following properties (see also Fig.\ 2):
\begin{enumerate}
\item $R_{\rm Sph}(y)$ is a non-negative, even, strictly convex function,
and $R_{\rm Sph}(0)=0$; \item $R_{\rm Sph}(y)\sim-\frac{1}{2}\ln(1-y^2)$, as
$y\to\pm 1$; \item $R''(0)=\frac{3-2\sqrt{2}}{\sqrt{2}}\beta J$,
for $\beta\geq\beta_c$.
\end{enumerate}

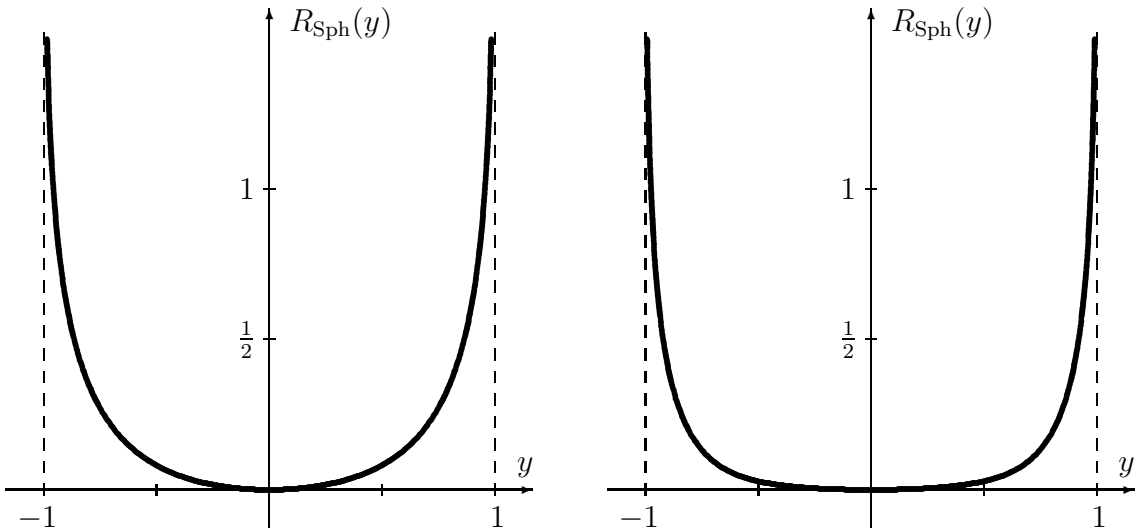
\begin{figure}
\setlength{\unitlength}{1mm}
\begin{picture}(150,70)(0,-1)
\put(0,5){\vector(1,0){70}} \put(35,0){\vector(0,1){69}}
\put(80,5){\vector(1,0){70}} \put(115,0){\vector(0,1){69}}
\put(69,8){\makebox(0,0){$y$}} \put(149,8){\makebox(0,0){$y$}}
\put(44.5,67){\makebox(0,0){$R_{\rm Sph}(y)$}}
\put(124.5,67){\makebox(0,0){$R_{\rm Sph}(y)$}}
\multiput(5,5)(0,3){21}{\makebox(0,0){$\rule{0.15mm}{1.5mm}$}}
\multiput(65,5)(0,3){21}{\makebox(0,0){$\rule{0.15mm}{1.5mm}$}}
\multiput(85,5)(0,3){21}{\makebox(0,0){$\rule{0.15mm}{1.5mm}$}}
\multiput(145,5)(0,3){21}{\makebox(0,0){$\rule{0.15mm}{1.5mm}$}}
\multiput(5,4.95)(15,0){5}{\makebox(0,0){$\rule{0.15mm}{1.5mm}$}}
\multiput(85,4.95)(15,0){5}{\makebox(0,0){$\rule{0.15mm}{1.5mm}$}}
\put(1.5,0){$-1$}\put(64.3,0){$1$}
\put(81.5,0){$-1$}\put(144.3,0){$1$}
\multiput(35,5)(0,20){3}{\makebox(0,0){$\rule{1.5mm}{0.15mm}$}}
\put(32,45){\makebox(0,0){$1$}}
\put(32,25){\makebox(0,0){$\frac{1}{2}$}}
\multiput(115,5)(0,20){3}{\makebox(0,0){$\rule{1.5mm}{0.15mm}$}}
\put(112,45){\makebox(0,0){$1$}}
\put(112,25){\makebox(0,0){$\frac{1}{2}$}}
\input cayley_rf_1
\input cayley_rf_2
\end{picture}
\caption{The rate functions for the magnetization of the spherical
model on a Cayley tree above ($\beta=\frac{1}{2}\beta_{\rm c}$,
left) and below ($\beta =2\beta_{\rm c}$, right) the critical
temperature. In both cases $R_{\rm Sph}(y)=0$ only for $y=0$, and
$R''_{\rm Sph}(y)>0$ for any $y\in(-1,1)$.}
\label{rsphpic}
\end{figure}

\section{Ground State Properties.}

It is always instructive to have a look at the ground-state
(zero-temperature) properties of the model under investigation. In
doing that we gain useful physical intuition and get an idea what
one could expect to find for non-zero temperatures.

It is shown in Appendix A that the unit-length eigenvectors
corresponding to the maximum eigenvalue of the Cayley-tree matrix are given by
\[
\{v_{l,m}^{(n,1)}\}_{(l,m)\in
T_n}=\pm\left\{\frac{2^{1-l/2}}{\sqrt{n+1}}\sin\frac{\pi
l}{n+1}\right\}_{(l,m)\in T_n}.
\]
Since configurations of the spherical model obey the constraint
\[
\sum_{(l,m)\in T_n}x_{l,m}^2=N\equiv 2^n - 1,
\]
the pair of ground-state configurations is given by
\[
\mbox{\boldmath $g$}_{\pm}=\pm\left\{\sqrt{\frac{2^{2-l}N}{n+1}}\sin\frac{\pi
l}{n+1}\right\}_{(l,m)\in T_n}.
\]
Now a simple calculation (see the calculation of $\gamma_{n,k}$
in Appendix B) shows that the large-$n$ asymptotics of the
ground-state magnetization is given by
\[
m_n=\pm\frac{2\pi}{(\sqrt{2}-1)^2}\,n^{-3/2}+O(n^{-5/2}),
\]
in the ``$+$" and the ``$-$" phase, respectively.
Thus, even below the critical temperature we should not expect to obtain non-zero
spontaneous magnetization in an infinite-tree limit. Instead we have to look at
the renormalized magnetization
\[
\rho_n=\frac{n^{3/2}}{N}\sum_{(l,m)\in T_n}x_{l,m},
\]
which, as it turns out, is the true order parameter for the spherical
model of a ferromagnet on a Cayley tree.

\section{Large Deviations of the Order Parameter.}

The results of zero-temperature analysis and the strict convexity of the rate function
$R_{\rm Sph}(y)$ below the critical temperature, see Eq.\ (\ref{lde1}) and Fig.\ \ref{rsphpic},
hint that our choice of the normalization for magnetization is not quite right.
That is, although the distribution of
\[
m_N=\frac{1}{N}\sum_{(j,k)\in T_n}x_{j,k}
\]
is indeed asymptotically degenerate and concentrates at 0, as
$N\to\infty$, but if we replace $N$ by a softer normalization,
then we could go back to the usual situation where the
distribution of order parameter concentrates at two points $\pm
\rho^*$. This is precisely the result that we are going to establish in
the present section.

For the distribution density of the renormalized magnetization
\[
r_{N,\gamma}=\frac{n^\gamma}{N}\sum_{(j,k)\in T_n}x_{j,k},
\]
where $\gamma>0$, we obtain
\begin{equation}
f_{N,\gamma}(\rho)=\frac{\beta JN}{2\pi
i\Theta_N}\left(\frac{\pi}{\beta J}
\right)^{(N-1)/2}\int\limits_{-i\infty+z_0}^{i\infty+z_0}\!\frac{dz}
{\sqrt{\Sigma_n(z)(z-\tau_{n,1})}}\exp\left[N\beta J
\Gamma_n(z,\rho)\right], \label{tmg}
\end{equation}
where
\[
\Gamma_n(z,\rho)=z-\frac{1}{2\beta J N}{\sum_{(j,k)\in T_n}}'
\ln(z-\lambda_{j,k})-\frac{N\rho^2}{n^{2\gamma}\Sigma_n(z)},
\]
and the prime indicates that the sum over $(j,k)$ does not include
the term $\ln(z-\tau_{n,1})$ corresponding to the (non-degenerate)
maximal eigenvalue. The multiplier $n^{2\gamma}$ suppresses the
blocking influence of $\Sigma_n(z)$ and allows the saddle-point of
$\Gamma_n(z,\rho)$ to enter the immediate vicinity of the
eigenvalue $\tau_{n,1}$ when $\beta\geq \beta_c$. But first we
have to find the distribution of $r_{N,\gamma}$ for high
temperatures.

If $\beta<\beta_c$, then the relevant saddle points $z_n(\rho)$ of
$\Gamma_n(z,\rho)$ do not approach $\tau_{n,1}$ and evaluation of
the integral (\ref{tmg}) is straightforward. The relevant solution
of the saddle-point equation
\[
1-\frac{L'(z)}{2\beta J}-\frac{\rho^2}{n^{2\gamma}}\sigma'(z)=0
\]
is given by
\[
z_n(\rho)=z^*-2\beta
J\rho^2\frac{\sigma'(z^*)}{L''(z^*)}n^{-2\gamma}+O(n^{-4\gamma}),
\]
where $z^*$ is the maximal solution of Eq.\ (\ref{lspe}).

Therefore, for $\beta<\beta_c$ the main asymptotics of the
distribution density of $r_{N,\gamma}$ is given by
\begin{equation}
f_{N,\gamma}(\rho)=\exp\left[-\frac{N}{n^{2\gamma}}\beta
J\sigma(z^*)\rho^2+O(Nn^{-4\gamma})\right]. \label{ldht}
\end{equation}
Note that the quadratic rate function $\beta J\sigma(z^*)\rho^2$
can be obtained formally from the rate function $R_{\rm Sph}(y)$,
see Eq.\ (\ref{lde1}), if we substitute $n^{-\gamma}\rho$ instead
of $y$. However,  below the critical temperature the situation
becomes very different.

To evaluate the integral in Eq.\ (\ref{tmg}) for $\beta>\beta_c$
we have to locate singularities of the integrand in the vicinity
of the maximal eigenvalue $z=\tau_{n,1}$. The singularities of
$L_n(z)$ and $\Sigma_n(z)$ at
\[
\tau_{n,1}=\sqrt{2}\left(1-\frac{\pi^2}{2}n^{-2}+\pi^2 n^{-3}\right)
+O(n^{-4})
\]
cancel each other, therefore the integrand is analytic at
$z=\tau_{n,1}$. The second-largest eigenvalue of the Cayley-tree matrix
\[
\tau_{n-1,1}=\sqrt{2}\left(1-\frac{\pi^2}{2}n^{-2}\right)+
O(n^{-4})
\]
is also non-degenerate and the point $z=\tau_{n-1,1}$ is the
sticking point --- the most important singularity of the
integrand. It prevents the saddle-point from sliding further to
the left when $\rho^2$ is small, and the eigenvector
$\mbox{\boldmath $v$}^{(n-1,1)}$, corresponding to $\tau_{n-1,1}$,
is the state absorbing the macroscopic ``condensation" taking
place when the temperature drops below a certain critical level
$T_{\rm ph.s.}$ where phase separation begins, see Eq.\
(\ref{betad}).

Another important singularity of the integrand is the maximal zero
of $\Sigma_n(z)$ at
\[
s_n=\sqrt{2}\left(1-\frac{\pi^2}{2}n^{-2}-2\pi^2(\sqrt{2}+1)n^{-3}\right)+
O(n^{-4}).
\]
It however lies to the left of $\tau_{n-1,1}$, a bit further away
from $\tau_{n,1}$, and, hence, in the present set up $s_n$ is not
the sticking point. That is, $z=s_n$ is not the singularity
preventing the saddle-point from sliding further to the left. The
main reason for that is orthogonality of the eigenvector
$\mbox{\boldmath $v$}^{(n-1,1)}$ and the constant vector
$x_{j,k}=1$ associated with the conventional and renormalized
magnetizations, $m_N$ and $r_{N,\gamma}$, respectively. However,
if we decide to look look at large-deviation probabilities of
other observables, for instance, of the magnetization of a
subdomain $D_n\subset T_n$, then the situation could become very
different, and the maximal zero of $\Sigma_n(z)$ could become the
sticking point.

The formulae for $\tau_{n,1}$, $\tau_{n-1,1}$, and $s_n$ suggest
that in order to evaluate the integral (\ref{tmg}) by the
saddle-point method we have to introduce a new integration
variable $\zeta$ via
\[
z=\sqrt{2}\left(1-\frac{\pi^2}{2}n^{-2}+ \pi^2\zeta n^{-3}\right).
\]
If $z>\tau_{n-1,1}$, that is, if $\zeta>0$, then the large-$n$ asymptotic expansion
of the function $\Gamma_n(z,\rho)$ is given by
\[
\Gamma_n\left[\sqrt{2}\left(1-\frac{\pi^2}{2n^{2}}+ \pi^2\zeta
n^{-3}\right),\rho\right]=\sqrt{2}\left(1-\frac{\pi^2}{2n^{2}}+
\pi^2\zeta n^{-3}\right)\left(1-\frac{\beta_c}{\beta}\right)
\]
\[
-\frac{1}{n^{2\gamma}2\sqrt{2}}\frac{(\sqrt{2}-1)^2(\zeta-1)\rho^2}
{(\sqrt{2}+1)^2+(\zeta-1)}+O(n^{-2(1+\gamma)}).
\]
The two $\zeta$-dependent terms are of the same magnitude if
$\gamma=\frac{3}{2}$. Differentiating the function $\Gamma_n(z,\rho)$
(with $\gamma$ replaced by $\frac{3}{2}$) over $\zeta$  we obtain
the following equation for the saddle point $\zeta^*$:
\[
\left(1-\frac{\beta_c}{\beta}\right)\pi^2-\frac{\rho^2}{4}\frac{1}
{\left((\sqrt{2}+1)^2+\zeta-1\right)^2}=0.
\]
Since $z=\tau_{n-1,1}$ is the sticking point, the relevant
solution $\zeta^*$ must be non-negative. Therefore
\[
\zeta^*=-2(1+\sqrt{2})+\frac{|\rho|}
{2\pi\sqrt{1-\frac{\beta_c}{\beta}}},
\]
if
\begin{equation}
|\rho|>\rho_c\equiv
4\pi(\sqrt{2}+1)\sqrt{1-\frac{\beta_c}{\beta}}, \label{muc}
\end{equation}
and $\zeta^*=0$, otherwise.

On applying the saddle-point method to the integral (\ref{tmg})
and using Eq.\ (\ref{maspflt}) for the main asymptotics of the
partition function we obtain the following expression for the
distribution density of the order parameter
\begin{equation}
f_{N,\frac{3}{2}}(\rho)=\exp\left(-Nn^{-3} R_{\rm
Cayley}(\rho)+O(N/n^{5})\right), \label{ldlt}
\end{equation} where
\[
R_{\rm Cayley}(\rho)=\sqrt{2}\beta J\times\left\{
\begin{array}{cc}
{\displaystyle
\left[\frac{\pi}{\sqrt{2}-1}\sqrt{1-\frac{\beta_c}{\beta}}-\frac{|\rho|}{2}(\sqrt{2}-1)
\right]^2},&\mbox{
if }|\rho|\geq\rho_c;\vspace{1mm}\\
{\displaystyle
\pi^2\left(1-\frac{\beta_c}{\beta}\right)-\frac{\rho^2}{8}(\sqrt{2}-1)^3},&\mbox{
if }|\rho|<\rho_c.
\end{array}
\right.
\]
The main features of the rate function $R_{\rm Cayley}(\rho)$ are shown in Figure 3.
It vanishes at the points
\[
\rho=\pm\rho^*\equiv\pm\frac{2\pi}{(\sqrt{2}-1)^2}\sqrt{1-\frac{\beta_c}{\beta}}.
\]
These points are the equilibrium values of the order parameter $r_{N,\frac{3}{2}}$
of the spherical model on a binary Cayley tree.

\begin{figure}
\setlength{\unitlength}{1mm}
\begin{picture}(150,70)(0,-1)
\put(0,5){\vector(1,0){70}} \put(35,0){\vector(0,1){69}}
\put(80,5){\vector(1,0){70}} \put(115,0){\vector(0,1){69}}
\put(68,8){\makebox(0,0){$\rho$}}
\put(148,8){\makebox(0,0){$\rho$}}
\put(45.5,67){\makebox(0,0){$R_{\rm Cayley}(\rho)$}}
\put(125.5,67){\makebox(0,0){$R_{\rm Cayley}(\rho)$}}
\multiput(5,5)(15,0){5}{\makebox(0,0){$\rule{0.15mm}{1.5mm}$}}
\put(95.579,5){\makebox(0,0){$\rule{0.15mm}{2mm}$}}
\put(134.421,5){\makebox(0,0){$\rule{0.15mm}{2mm}$}}
\multiput(131.089,5)(0,3){10}{\makebox(0,0){$\rule{0.15mm}{1.5mm}$}}
\multiput(98.911,5)(0,3){10}{\makebox(0,0){$\rule{0.15mm}{1.5mm}$}}
\put(91,0){$-\rho^*$} \put(133,0){$\rho^*$}
\put(104,8){\makebox(0,0){$-\rho_c$}}
\put(128,8){\makebox(0,0){$\rho_c$}}
\put(14.5,0){$-10$}\put(48,0){$10$}
\multiput(35,5)(0,20){4}{\makebox(0,0){$\rule{1.5mm}{0.15mm}$}}
\put(31,45){\makebox(0,0){$10$}}
\multiput(115,5)(0,20){4}{\makebox(0,0){$\rule{1.5mm}{0.15mm}$}}
\put(111,45){\makebox(0,0){$10$}}
\input cayley_rf_3
\input cayley_rf_4
\end{picture}
\caption{Rate functions for the order parameter
$r_{N,\frac{3}{2}}$ of the spherical model on a Cayley tree for
$\beta=\beta_{\rm c}$ (left) and for $\beta =2\beta_{\rm c}$
(right).}
\end{figure}
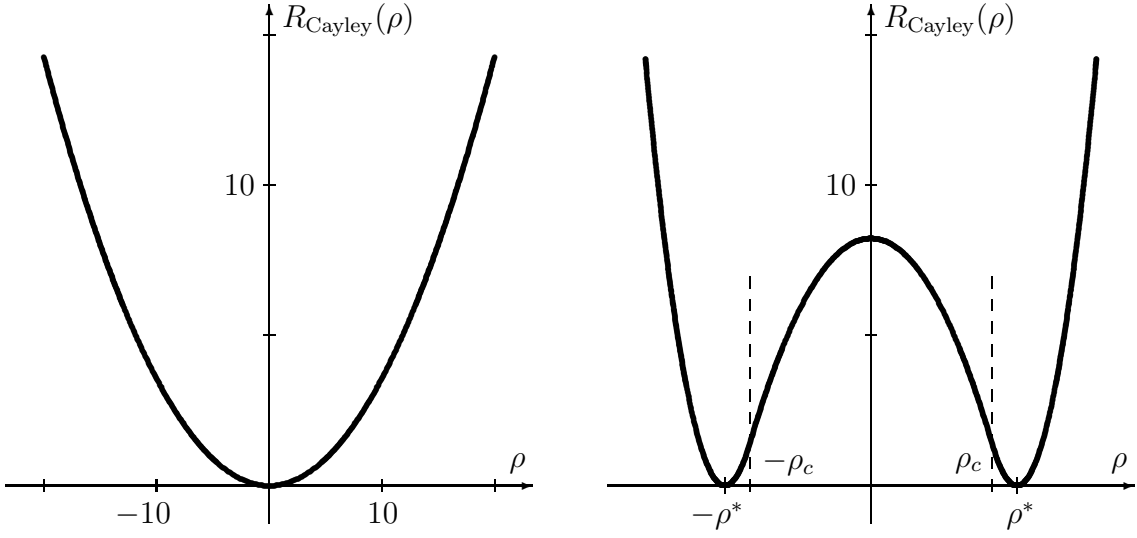

The critical level $\rho_c$, given by Eq. (\ref{muc}), can be also interpreted as a
relationship for the critical temperature of the ``spherical" lattice gas on a Cayley tree.
If we consider the ensemble with the gas density
\[
\frac{1}{N}\sum_{(j,k)\in T_n}x_{j,k}
\]
fixed at $n^{-3/2}\rho$, then, as it follows from Eq.\
(\ref{muc}), a phase separation takes place in the system when the
temperature drops below
\begin{equation}
T_{\rm ph.s}=T_{\rm
c}\left[1-\left(\frac{\sqrt{2}-1}{4\pi}\right)^2\rho^2\right].
\label{betad}
\end{equation}
Since the saddle-point of the integral (\ref{tmg}) sticks at the
second-largest eigenvalue $\tau_{n-1,1}$, the formula for the
eigenvector $\mbox{\boldmath $v$}^{(n-1,1)}$ suggest the following
picture below the phase-separation temperature $T_{\rm ph.s}$. The
high-density phase ($x_{j,k}>0$) gathers in one half of the binary
tree, while the low-density phase ($x_{j,k}<0$) is all that
remains in the other half of the binary tree.

\section{Discussion and concluding remarks.} The analysis of
previous sections can be extended to the case of a Cayley tree
with branching ratio $q>2$ at the expense of extra technical
efforts. According to \cite{bdp_1992}, the second-largest
eigenvalue $\tau_{n-1,1}$ is $(q-1)$-times degenerate. This
degeneracy has certain macroscopically observable consequences in
lattice-gas models, that is, in models with fixed value of
properly normalized gas density (total spin).

Phase separation in lattice-gas spherical models is a condensation
into eigenvectors corresponding to the second-largest eigenvalue
$\tau_{n-1,1}$. In the case $q=2$ the condensation scenario is
quite simple. An $n$-generation Cayley tree, $T_n$, consists from
the root, the left sub-tree $T_{n-1}^{\rm L}$ and the right
subtree $T_{n-1}^{\rm R}$, see Fig.\ \ref{np1}. Accordingly, the
eigenvector $\mbox{\boldmath $v$}^{(n-1,1)}$ is a combination of
the high-density and low-density ground states in the left and
right subtrees (or vice-versa) and zero-value at the root.
Therefore, the condensation scenario on a binary tree is quite
simple. If the fixed value of normalized gas density is not equal
to the equilibrium value, then the excess of the high-density
phase gathers in the left or right subtree, while the low-density
phase gathers in the opposite subtree.

If we consider a Cayley tree with branching ratio $q>2$, then the
number of phase-separation scenarios increases. For instance, if
$q=3$, then $T_n$ contains three $(n-1)$-generation subtrees, and
the high-density phase could gather either in one or in two of the
three available subtrees.

The widely known solution of the Ising model on a Cayley tree (the
IC model) described in the book by Baxter \cite{b_1982} is based
on calculating the magnetization induced by a homogeneous field
$h$ applied at the boundary. If the temperature is sufficiently
high, then (in the thermodynamic limit) the boundary field has no
influence on the random variables $x_{j,k}$ located close to the
root of the tree. However, when the temperate is below $T_{\rm
B}$: $\tanh(J \beta_{\rm B})=1/2$, an arbitrarily weak boundary
field $h$ induces non-zero expected values of all random
variables, including those located around the root of the tree.
Namely, the boundary field induces a magnetization $m_N(h)$ in the
middle of the tree, and $m_N(h)$ converges to a non-zero limit
$m(T)\,{\rm sgn}(h)$ as $N\to\infty$. Moreover, $m(T)$ does not
depends on $h$. The penetration temperature $T_{\rm B}$ was
interpreted in \cite{b_1982} as the critical temperature of the IC
model.

As a rule, the penetration temperature and the critical
temperature coincide in finite-dimensional systems. However, the
situation becomes very different when we consider models on Cayley
trees. Indeed, the exact solution reported in the present paper
shows that the critical and the penetration temperatures differ in
the case of the spherical model. For the characteristic function
\[
\chi_{j,k}(t)=\langle\exp(i t x_{j,k})\rangle_n
\]
of the random variable $x_{j,k}$ at the node $(j,k)\in T_n$ we
obtain the following large-$n$ asymptotics
\begin{equation} \chi_{j,k}(t)\sim
\exp\left[-\frac{t^2}{4\beta J}\sum_{(l,m)\in
T_n}\frac{\left(v_{j,k}^{(l,m)}\right)^2}{z_n^*-\lambda_{l,m}}+\frac{it}{2
J}\sum_{(l,m)\in T_n}\frac{\phi_{l,m} v_{j,k}^{(l,m)}}{z_n^*
-\lambda_{l,m}}\right], \label{cfid}
\end{equation}
where $\phi_{l,m}$ are the scalar products of the eigenvectors
$\mbox{\boldmath $v$}^{(l,m)}$ and the homogeneous boundary field,
see Eq.\ (\ref{hphi}). Since $\phi_{l,m}\neq 0$ only for $l=n$,
the large-$n$ asymptotics of the expected values of the random
variables $x_{j,k}$ are given by
\[
\langle x_{j,k}\rangle_n=\frac{1}{2 J}\sum_{(l,m)\in
T_n}\frac{\phi_{l,m} v_{j,k}^{(l,m)}}{z_n^* -\lambda_{l,m}}=
\frac{2^{(n-j)/2}h}{2(n+1)J}\sum_{m=1}^n\frac{2\sin\frac{\pi n
m}{n+1}\sin\frac{\pi j m}{n+1}}{z_n^*-\sqrt{2}\cos\frac{\pi
m}{n+1}}.
\]
Equation (\ref{us1}) yields
\[
\langle x_{j,k}\rangle_n= \frac{2^{(n-j+1)/2}h}{2J}\frac{
x_+^{j}(z_n^*)-x_-^{j}(z_n^*)}
{x_+^{n+1}(z_n^*)-x_-^{n+1}(z_n^*)}.
\]
Therefore, the effective field generated by the boundary conditions penetrates
inside the tree once $x_+(z^*)\leq\sqrt{2}$, that is, once
$z^*\leq\frac{3}{2}$. Looking at the saddle-point equation (\ref{spl3}) we
conclude that the penetration temperature of the spherical model
is given by
\[
T_{\rm p}=\frac{J}{W_{\rm
Cayley}(3/2)}\left(1-\frac{1}{\sqrt{2}}\left(\frac{h}{2J}\right)^2
\right),
\]
where
\[
W_{\rm Cayley}(3/2)=\sum_{j=2}^\infty
2^{-j}\left(j\frac{2^j+1}{2^j-1}-3\right).
\]
The penetration temperature $T_{\rm p}$ is the direct analogue of the critical
temperature of the Ising model on a Bethe lattice, $T_{\rm B}$.

The large-deviation probabilities for the order-parameter
$r_{n,\frac{3}{2}}$ decay exponentially with $N/n^3$ (not with
$N$). Nevertheless, the low-temperature phases of the spherical
model on a Cayley tree should be classified as rigid. Indeed, the
benchmark of rigidity is the behavior of large-deviation
probabilities for $T>T_c$. According to Eqs.\ (\ref{ldht}) and
(\ref{ldlt}) large-deviation probabilities for $r_{n,\frac{3}{2}}$
decay exponentially with $N/n^3$ both below and above the critical
temperature $T_c$. Therefore, the low-temperature phases are
rigid.

\newpage

\noindent {\bf\large Appendix A. Spectral Properties of
Cayley-Tree Matrices. 
} \vspace{\abovedisplayskip}

Often, methods developed for solving various 1D models are
successfully applied to the corresponding models on Cayley trees.
Not surprisingly, after a minor effort, calculation of eigenvalues
$\lambda_{k,l}$ and eigenvectors $\mbox{\boldmath $u$}^{(k,l)}$ of
the Cayley-tree matrix $\widehat M_N$ having $N=2^n-1$ rows and
columns, see Eq.\ (\ref{ctm}), is reduced to investigation of
spectral properties of tri-diagonal matrices.

We use the symbols $\lambda_{k,l}$ and $\mbox{\boldmath
$u$}^{(k,l)}$ to denote an abstract complete set of $N$
eigenvalues and orthonormal eigenvectors of the Cayley-tree matrix
$\widehat M_N$. For instance, we use these notations to
diagonalise the Hamiltonian $H_n$ of the spherical model, see Eq.\
(\ref{ham}). In this case, the ranges of indexes $k$ and $l$ in
the eigenvalues $\lambda_{k,l}$ and eigenvectors $\mbox{\boldmath
$u$}^{(k,l)}$ mimic the labelling of nodes in the tree $T_n$:
$l=1,2,\ldots,2^{k-1}$; and $k=1,2,\ldots,n$.

We use the symbols $\tau_{k,l}$ and $\mbox{\boldmath
$v$}^{(k,l;j)}$ to denote the spectrum (the set of different
eigenvalues) of the matrix $\widehat M_N$ and the corresponding
eigenvectors. In this case, as we shall see below, the indexes
$k$ and $l$ run over the triangular array $l=1,2,\ldots,k$;
$k=1,2,\ldots,n$. The multiplicities of eigenvalues $\tau_{k,l}$
are denoted either $m_{n;k,l}$, or $m_{k,l}$ if it is clear from
the context how many generation the Cayley tree contains. The last
index in $\mbox{\boldmath $v$}^{(k,l;j)}$, separated from the
others by a semicolon, reflects the multiplicity of eigenvalue
$\tau_{k,l}$: $j=1,2,\ldots,m_{k,l}$. If $m_{k,l}=1$, then one can
omit the multiplicity index $j$, $\mbox{\boldmath
$v$}^{(k,l;j)}=\mbox{\boldmath $v$}^{(k,l;1)}=\mbox{\boldmath
$v$}^{(k,l)}$. In the recursive procedure used below for
constructing the complete set of eigenvectors of $\widehat M_N$ it
might be necessary to specify explicitly the number of tree
generations, $n$. Alas, in such cases we have to append one more
superscript to eigenvectors and use symbols like $\mbox{\boldmath
$v$}^{(n;k,l;j)}$ or $\mbox{\boldmath $v$}^{(n;k,l)}$. If
specified at all, the number of generations, $n$,  is always the
first single index separated from the others by a semicolon.

We use two kinds of indexing for the components of a vector
$\mbox{\boldmath $x$}$. When the components must be ordered explicitly
we use a single index, $\mbox{\boldmath $x$}=\{x_j:
j=1,2,...,2^n-1\}$. If it is necessary to take into account the
tree structure, we mimic the indexing of nodes in the tree $T_n$:
$\mbox{\boldmath $x$}=\{x_{i,j}: j=1,2,...,2^{i-1};
i=1,2,...,n\}$. The component $x_j$ with $j=2^{n-1}$ of
an explicitly ordered vector $\mbox{\boldmath $x$}$ always
corresponds to the root (the node $(1,1)$) of the tree $T_n$.
Therefore, we call $x_j$ with $j=2^{n-1}$ the root component.

We begin our quest for spectral properties of Cayley-tree matrix
$\widehat M_N$ with finding all special eigenvectors
$\mbox{\boldmath $v$}\equiv\{v_{j,k}:(j,k)\in T_n\}$ having the
following form
\begin{equation}
v_{j,1}=v_{j,2}=\ldots =v_{j,2^{j-1}}=y_j,\quad\mbox{for
}j=1,2,\ldots,n. \label{sev1}
\end{equation}
That is, our first aim is to find all eigenvectors with identical
components along each generation of the tree. Since, the
components $v_{j,k}$ of any eigenvector satisfy a linear
relationship of the form
\[
{\textstyle\frac{1}{2}}v_{j-1,l}+{\textstyle\frac{1}{2}}v_{j+1,m}+
{\textstyle\frac{1}{2}}v_{j+1,m+1}=\lambda v_{j,k},
\]
the components of the vector $\mbox{\boldmath
$y$}\equiv\{y_j\}_{j=1}^n$ satisfy the relationship
$\frac{1}{2}y_{j-1}+y_{j+1}=\lambda y_j$. Hence, the vector
$\mbox{\boldmath $y$}$ is one of the eigenvectors of the $n\times n$ tri-diagonal matrix
\[
\widehat L_n=\left(
\begin{array}{ccccccc}0&1&&&&&\\
\frac{1}{2}&0&1&&&\mbox{\LARGE 0}&\\
&\frac{1}{2}&0&\ddots&&&\\
&&\ddots&\ddots&\ddots&&\\
&&&\ddots&0&1&\\
&\mbox{\LARGE 0}&&&\frac{1}{2}&0&1\\
&&&&&\frac{1}{2}&0
\end{array}
\right).
\]
Therefore (non-degenerate) eigenvalues $\tau_{n,l}$
corresponding to
special eigenvectors $\mbox{\boldmath $v$}^{(n,l)}$ of the matrix $\widehat M_N$ coincide with the
eigenvalues $\Lambda_{n;l}=\sqrt{2}\cos\frac{\pi l}{n+1}$, $l=1,2,\ldots,n$ of the matrix $\widehat L_n$.
The eigenvectors of $\widehat L_n$ are given by
\[
\mbox{\boldmath
$y$}^{(n;l)}=\left\{y_k^{(n;l)}\right\}_{k=1}^n=\left\{
2^{-k/2}\sin\frac{\pi kl}{n+1}\right\}_{k=1}^n,\quad
k=1,2,\ldots,n.
\]
Note that the first component $y_1^{(n;l)}$ of any eigenvector
$\mbox{\boldmath $y$}^{(n;l)}$ is greater than $0$.

Since the matrix $\widehat L_n$ is asymmetric, the vectors
$\mbox{\boldmath $y$}^{(n;l)}$ are linearly independent but they
are not orthogonal. Formally, the reason for the lack of
orthogonality are the multipliers $2^{-k/2}$. However, there are
exactly $2^{k-1}$ nodes $(k,j)$ in the $k$-th generation of a
binary tree $T_n$. Therefore, the special eigenvectors
$\mbox{\boldmath $v$}^{(n,l)}$ constructed from the vectors
$\mbox{\boldmath $y$}^{(n;l)}$ according to Eq.\ (\ref{sev1}) are
orthogonal.

Complete sets of eigenvalues and
eigenvectors for the matrices
\[
\widehat M_N\quad \mbox{with}\quad N=2^n-1,\quad n=2,3,\ldots
\]
can be constructed from the eigenvalues and eigenvectors of
$\widehat M_{(N-1)/2}$ and from the eigenvalues
$\Lambda_{n;l}(=\tau_{n,l})$ of the matrix $\widehat L_n$ and the
corresponding special eigenvectors $\mbox{\boldmath $v$}^{(n,l)}$.
In the case of a tree with two generations, $T_2$, the eigenvalues
and eigenvectors of the matrix $\widehat L_2$ are given by
\[
\Lambda_{2;1}=-\frac{1}{\sqrt{2}},\quad\Lambda_{2;2}=\frac{1}{\sqrt{2}}\quad \mbox{and}
\qquad\mbox{\boldmath $y$}^{(2;1)}=\left(
\begin{array}{r}
\sqrt{2}\\ -1\end{array}
\right), \quad\mbox{\boldmath $y$}^{(2;2)}= \left(
\begin{array}{r}
\sqrt{2}\\ 1
\end{array}
\right).
\]
The corresponding special eigenvectors of the
Cayley-tree matrix
\begin{equation} \widehat M_3= \left(
\begin{array}{ccc}
0& \frac{1}{2}& 0\\ \frac{1}{2}& 0 &\frac{1}{2}\\ 0& \frac{1}{2}& 0
\end{array}
\right) \  \mbox{are given by}\ \ \mbox{\boldmath
$v$}^{(2,1)}=\left(
\begin{array}{r}
-1\\ \sqrt{2}\\ -1
\end{array}
\right), \ \mbox{\boldmath $v$}^{(2,2)}= \left(
\begin{array}{c}
1\\ \sqrt{2}\\ 1
\end{array}
\right). \label{evm3}
\end{equation}
The remaining eigenvalue of the matrix $\widehat M_3$ is
$\tau_{1,1}=0$, and the corresponding eigenvector is
$\mbox{\boldmath $v$}^{(1,1)}=\left(1,0,-1\right)^{\rm T}$.

Before turning to the general induction step it is instructive to
see how one can construct eigenvalues and eigenvectors of the
Cayley-tree matrix
\[
\widehat M_{7}=\left(
\begin{array}{ccccccc}0&0&\frac{1}{2}&&&&\\
0&0&\frac{1}{2}&&&\mbox{\LARGE 0}&\\
\frac{1}{2}&\frac{1}{2}&0&\frac{1}{2}&&&\\
&&\frac{1}{2}&0&\frac{1}{2}&&\\
&&&\frac{1}{2}&0&\frac{1}{2}&\frac{1}{2}\\
&\mbox{\LARGE 0}&&&\frac{1}{2}&0&0\\
&&&&\frac{1}{2}&0&0
\end{array}
\right),
\]
from the eigenvalues and eigenvectors of the matrix $\widehat M_3$.

The last three eigenvalues of $\widehat M_7$ are the eigenvalues
of the matrix $\widehat L_3$:
\[
\tau_{3,1}=\Lambda_{3;1}=-1,\quad \tau_{3,2}=\Lambda_{3;2}=0,\quad \tau_{3,3}=\Lambda_{3;3}=1.
\]
The eigenvectors of the matrix $\widehat L_3$ are given by
\[
\mbox{\boldmath $y$}^{(3;1)}=\left(
\begin{array}{r}
1\\ -1\\ \frac{1}{2}
\end{array}
\right), \qquad\mbox{\boldmath $y$}^{(3;2)}= \left(
\begin{array}{r}
1\\ 0\\ -\frac{1}{2}
\end{array}
\right),  \qquad\mbox{\boldmath $y$}^{(3;3)}= \left(
\begin{array}{r}
1\\ 1\\ \frac{1}{2}
\end{array}
\right).
\]
Hence, the special (constant along generation)
eigenvectors of the matrix $\widehat M_7$ are
\begin{eqnarray}
&&\mbox{\boldmath $v$}^{(3,1)}=\left(\frac{1}{2},\frac{1}{2},
-1,1,-1,\frac{1}{2},\frac{1}{2}\right)^{\rm T},\nonumber \\
&& \mbox{\boldmath
$v$}^{(3,2)}=\left(-\frac{1}{2},-\frac{1}{2},0,1,0,-\frac{1}{2},-\frac{1}{2}\right)^{\rm
T},\nonumber  \\ &&\mbox{\boldmath
$v$}^{(3,3)}=\left(\frac{1}{2},\frac{1}{2},1,1,1,\frac{1}{2},\frac{1}{2}\right)^{\rm
T}.\nonumber
\end{eqnarray}

The remaining eigenvalues $\tau_{1,1}$, $\tau_{2,1}$, $\tau_{2,2}$ of $\widehat M_7$ are
identical to the eigenvalues $\tau_{k,l}$ of the matrix
$\widehat M_3$, but, as we will see shortly, the eigenvalue
$\tau_{1,1}$ of $\widehat M_7$ is twice degenerate, $m_{3;1,1}=2$.

To find the eigenvectors of the matrix $\widehat M_7$
corresponding to $\tau_{1,1}$, take the eigenvector
$\mbox{\boldmath $v$}^{(2;1,1)}$ of $\widehat M_3$ and note that
the root component of $\mbox{\boldmath $v$}^{(2;1,1)}$ is zero,
$v_2^{(2;1,1)}=0$. The vectors $(1,-1,0)^{\rm T}$ and
$(0,-1,1)^{\rm T}$ --- permutations of $\mbox{\boldmath
$x$}^{(2;1,1)}$ --- are eigenvectors of the $3\times 3$ blocks in,
respectively, the upper left and lower right corners of $\widehat
M_7$. Therefore the vectors
\[
\mbox{\boldmath $v$}^{(3;1,1;1)}=\left(1,-1,0,0,0,0,0\right)^{\rm T}\quad \mbox{and}\quad
\mbox{\boldmath $v$}^{(3;1,1;2)}=\left(0,0,0,0,0,-1,1\right)^{\rm T}
\]
are eigenvectors of the matrix $\widehat M_7$ corresponding to the
eigenvalue $\tau_{1,1}$. Note that we are able to construct two
orthogonal eigenvectors of the matrix $\widehat M_7$ from the
eigenvector $\mbox{\boldmath $v$}^{(2;1,1)}$, because the root
component of $\mbox{\boldmath $v$}^{(2;1,1)}$ is equal to 0.

The remaining eigenvectors of $\widehat M_7$ are constructed
from the eigenvectors $\mbox{\boldmath $v$}^{(2;2,1)}$ and $\mbox{\boldmath $v$}^{(2;2,2)}$ of
the matrix $\widehat M_3$, see Eq.\ (\ref{evm3}).

The permutations
\[
\left(
\begin{array}{r}
1\\ 1\\ \sqrt{2}
\end{array}
\right) \qquad\mbox{and}\qquad\left(
\begin{array}{r}
\sqrt{2}\\ 1\\ 1
\end{array}
\right)
\]
of $\mbox{\boldmath $v$}^{(2;2,2)}$ are eigenvectors of,
respectively, the upper left and lower right $3\times 3$ blocks of
the matrix $\widehat M_7$. An inspection shows that the
anti-symmetric vector
\[
\mbox{\boldmath
$v$}^{(3;2,2)}=\left(1,1,\sqrt{2},0,-\sqrt{2},-1,-1\right)^{\rm T}
\]
is the eigenvector of the matrix $\widehat M_7$ corresponding to
the eigenvalue $\lambda_{2,2}=\frac{1}{\sqrt{2}}$. Analogously,
the anti-symmetric vector
\[
\mbox{\boldmath
$v$}^{(3;2,1)}=\left(-1,-1,\sqrt{2},0,-\sqrt{2},1,1\right)^{\rm T}
\]
constructed from permutations of $\mbox{\boldmath $v$}^{(2;2,1)}$
is the eigenvector of the matrix $\widehat M_7$ corresponding to
the eigenvalue $\lambda_{2,1}=-\frac{1}{\sqrt{2}}$. Note that the
root components of the eigenvectors $\mbox{\boldmath
$v$}^{(2;2,1)}$ and $\mbox{\boldmath $v$}^{(2;2,2)}$ are not equal
to zero and each of those vectors generates exactly one
eigenvector of the matrix $\widehat M_7$. The three special
eigenvectors together with the four eigenvectors constructed from
the eigenvectors of the matrix $\widehat M_3$ make up a complete
set of eigenvectors of the matrix  $\widehat M_7$.

Using essentially the same procedure one can construct the
eigenvalues and eigenvectors of a matrix $\widehat M_{2N+1}$ from
those of the matrix $\widehat M_{N}$, where $N=2^n-1$. The first
$n(n+1)/2$ eigenvalues $\tau_{k,l}$ of $\widehat M_{2N+1}$ are the
eigenvalues of $\widehat M_{N}$. The remaining $n+1$ eigenvalues
of $\widehat M_{2N+1}$ are the eigenvalues $\tau_{n+1,l}=
\Lambda_{n+1;l}$ of the matrix $\widehat L_{n+1}$. A complete set
of $n+1$ eigenvectors of $\widehat L_{n+1}$ generates $n+1$
special (constant along generations) orthogonal eigenvectors of
the matrix $\widehat M_{2N+1}$. Each eigenvector of $\widehat
M_{N}$ with the root component equal to 0 generates a pair of
orthogonal eigenvectors of $\widehat M_{2N+1}$ (with zero root
components). Finally, the set of $n$ special eigenvectors of
$\widehat M_{N}$ generates $n$ orthogonal anti-symmetric
eigenvectors of $\widehat M_{2N+1}$ (with the root components
equal to zero). Note that according to the above construction only
special eigenvectors have non-zero root components.

\begin{figure}[t]
\setlength{\unitlength}{1mm}
\begin{picture}(150,70)(0,5)

\put(80,54){ $x_{1,1}$} \put(77,54){{\circle*{1}}}

\put(15,30){\Large $T_{n+1}$:}

\multiput(77.2,54)(1.2,-0.8){21}{\makebox(0,0){\circle*{0.2}}}
\multiput(77.2,54)(-1.2,-0.8){21}{\makebox(0,0){\circle*{0.2}}}

\put(44.5,38){$x_{2,1}$} \put(103.5,38){$x_{2,2}$}
\put(53,38){{\circle*{1}}} \put(101,38){{\circle*{1}}}

\put(50,18){\Large $T_{n}^{\rm L}$} \put(98,18){\Large $T_{n}^{\rm R}$}

\thicklines

\put(53,38){\line(1,-2){15}} \put(53,38){\line(-1,-2){15}}
\put(101,38){\line(1,-2){15}} \put(101,38){\line(-1,-2){15}}
\put(38,8){\line(1,0){30}} \put(86,8){\line(1,0){30}}

\end{picture}
\caption{An $(n+1)$-generation Cayley tree, $T_{n+1}$, consists
from the root, with the attached variable $x_{1,1}$, and from two
identical $n$-generation trees, $T_{n}^{\rm L}$ and $T_{n}^{\rm
R}$ (left and right), with the variables $x_{2,1}$ and $x_{2,2}$
attached to their roots.} \label{np1}
\end{figure}

To justify the above recursive procedure we construct the matrix
$\widehat M_{2N+1}$ from a pair of matrices $\widehat M_{N}$ used
as building blocks. We begin from arranging in a simple sequence
the labels $(j,k)$ of the matrix elements $M_{(j,k),(l,m)}$. That
is, we number the labels $(j,k)$ by integers $1,2,\ldots,2N+1$.
Denote $\widehat U_N$ the matrix obtained from  $\widehat M_{N}$
if we number the labels from the bottom of a tree to the top. The
exact numbering algorithm is not important, but the root $(1,1)$
must receive the highest number $N\equiv 2^n-1$. Denote $\widehat
D_N$ the matrix obtained by numbering the labels $(j,k)$ of
$\widehat M_{N}$ in the opposite order, from top to bottom. Now
$M_{(1,1),(k,l)}$ are the elements of the first row of the matrix
$\widehat M_N$. Note that, if $(v_1,v_2,\ldots,v_N)$ is an
eigenvector of $\widehat U_N$, then $(v_N,v_{N-1},\ldots,v_1)$ is
an eigenvector of $\widehat D_N$ with the same eigenvalue.

Using the two matrices $\widehat U_N$ and $\widehat D_N$ as
building blocks we can construct the Cayley-tree matrix $M_{2N+1}$
as shown in Fig.\ \ref{np1}. We number the labels of the left
subtree by integers $1,2,\ldots,2^n-1$ from bottom to top, assign
the number $2^n$ to the root of $T_{n+1}$, and number the right
subtree by integers $2^n+1,2^n+2,\ldots,2^{n+1}-1$ from top to
bottom in the opposite order. The obtained Cayley-tree matrix
$\widehat M_{2N+1}$ looks like this
\[
\widehat M_{2N+1}=\left(
\begin{array}{ccccccc}u_{1,1}&\ldots&u_{1,N}&&&&\\
\vdots&\ddots&\vdots&&&\mbox{\LARGE 0}&\\
u_{N,1}&\ldots&u_{N,N}&\frac{1}{2}&&&\\
&&\frac{1}{2}&0&\frac{1}{2}&&\\
&&&\frac{1}{2}&d_{1,1}&\ldots&d_{1,N}\\
&\mbox{\LARGE 0}&&&\vdots&\ddots&\vdots\\
&&&&d_{N,1}&\ldots&d_{N,N}
\end{array}
\right).
\]
If we already know the eigenvalues and eigenvectors of the
matrices $\widehat U_N$ and $\widehat D_N$, then the eigenvectors
of the matrix $\widehat M_{2N+1}$ are constructed as follows. If
$\mbox{\boldmath $v$}$ is an eigenvector of $\widehat U_N$ with
eigenvalue $\tau$ and with zero root component, that is, if
$v_N = 0$, then $(v_1,\ldots,v_N,0,\ldots,0)$ is an eigenvector of
$\widehat M_{2N+1}$ with the same eigenvalue $\tau$.
Due to the simple relationship between the eigenvectors of
$\widehat U_N$ and $\widehat D_N$ mentioned above, the vector
$(0,\ldots,0,v_N,\ldots,v_1)$ is also an eigenvector of $\widehat
M_{2N+1}$ with the eigenvalue $\tau$.

Let now $\mbox{\boldmath $v$}^{(n;n,l)}$ be a special eigenvector
of $\widehat U_N$ with the eigenvalue $\tau_{n,l}$,  then
\[
\mbox{\boldmath $v$}^{(n+1;n,l)}\equiv
(v_1,\ldots,v_N,0,-v_N,\ldots,-v_1)
\]
is an eigenvector of $\widehat M_{2N+1}$ with the same eigenvalue.
Note that $\mbox{\boldmath $v$}^{(n+1;n,l)}$ is an anti-symmetric
vector with zero root component.

Now we are going to show that the obtained eigenvectors with
zero-root components and the $n+1$ special eigenvectors obtained
from linearly independent eigenvectors of the matrix $\widehat
L_{n+1}$ comprise the complete set of orthogonal eigenvectors of
the matrix $\widehat M_{2N+1}$. Since the constructed eigenvectors
are orthogonal, the obtained set is complete if it contains $2N+1$
vectors. According to our recursive procedure, the matrix
$\widehat M_{N}$ has $n$ special eigenvectors and $N-n$
eigenvectors with zero root component. Therefore, the matrix
$\widehat M_{2N+1}$, has exactly $2(N-n)+n=2N-n$ eigenvectors with
zero root component. Together with $n+1$ special eigenvectors we
obtain $2N+1$ orthogonal vectors. Therefore the obtained set is
the complete set of eigenvectors of the matrix $\widehat
M_{2N+1}$.
\newpage

\noindent {\bf\large Appendix B. Useful Sums and their
Asymptotics.} \vspace{\abovedisplayskip}

In this appendix we derive an explicit expression for the sum
\[
\Sigma_n(z)=\sum_{(j,k)\in
T_n}\frac{\gamma_{j,k}^2}{z-\lambda_{j,k}},
\]
see Eq.\ (\ref{sgm}), but we begin with a formula for the coefficients
\[
\gamma_{j,k}\equiv\sum_{(l,m)\in T_n}u_{l,m}^{(j,k)}.
\]

By construction, see Appendix A, only the special (constant along
generations) eigenvectors $\mbox{\boldmath $v$}^{(n,k)}$ have
non-zero sums of their components. Therefore
\[
\Sigma_n(z)=\sum_{k=1}^n\frac{\delta_{n,k}^2}{z-\tau_{n,k}},\quad
\mbox{where}\quad\delta_{n,k}=\sum_{(l,m)\in T_n}v_{l,m}^{(n,k)}.
\]

The normalized special eigenvectors $\mbox{\boldmath $v$}^{(n,k)}$, $k=1,2,\ldots,n$,
are given by
\[
\left\{v_{l,m}^{(n,k)}\right\}_{(l,m)\in
T_n}=\left\{\frac{2^{1-l/2}}{\sqrt{n+1}}\sin\frac{\pi
lk}{n+1}\right\}_{(l,m)\in T_n}.
\]
Hence
\begin{eqnarray}
\delta_{n,k}&=&\sum_{l=1}^n\sum_{m=1}^{2^{l-1}}\frac{2^{1-l/2}}{\sqrt{n+1}}\sin\frac{\pi
l k}{n+1}=\frac{1}{\sqrt{n+1}}\sum_{l=1}^n 2^{l/2}\sin\frac{\pi
l k}{n+1}\nonumber\\
&=&\frac{1}{\sqrt{n+1}}\,\mbox{Im}\sum_{l=1}^n
\left(\sqrt{2}\exp\frac{i\pi
k}{n+1}\right)^l=\sqrt{\frac{2}{n+1}}\sin\frac{\pi
k}{n+1}\frac{1-(-1)^k 2^{\frac{n+1}{2}}}{3-2\sqrt{2}\cos\frac{\pi
k}{n+1}}.\nonumber
\end{eqnarray}
Thus
\[
\Sigma_n(z)=\frac{2}{n+1}\sum_{k=1}^n\frac{\sin^2\frac{\pi
k}{n+1}}{(3-2\sqrt{2}\cos\frac{\pi
k}{n+1})^2}\frac{1+2^{n+1}-(-1)^k
2^{\frac{n+3}{2}}}{z-\sqrt{2}\cos\frac{\pi k}{n+1}}.
\]

To find a manageable expression for $\Sigma(z)$ let us introduce
an extra parameter $a$ and consider the sum
\[
S(a,z)=\frac{2}{n+1}\sum_{k=1}^n\frac{\sin^2\frac{\pi
k}{n+1}}{(a-2\sqrt{2}\cos\frac{\pi
k}{n+1})^2}\frac{1}{z-\sqrt{2}\cos\frac{\pi k}{n+1}}.
\]
Now one can decimate the degree of the denominator and split the
sum in two parts
\begin{eqnarray}
S(a,z)&=&-\frac{2}{n+1}\frac{d}{da}\sum_{k=1}^n\frac{\sin^2\frac{\pi
k}{n+1}}{a-2\sqrt{2}\cos\frac{\pi
k}{n+1}}\frac{1}{z-\sqrt{2}\cos\frac{\pi k}{n+1}}\nonumber\\
&=&-\frac{1}{n+1}\frac{d}{da}\frac{1}{z-a/2}
\left(\sum_{k=1}^n\frac{\sin^2\frac{\pi
k}{n+1}}{a/2-\sqrt{2}\cos\frac{\pi
k}{n+1}}-\sum_{k=1}^n\frac{\sin^2\frac{\pi
k}{n+1}}{z-\sqrt{2}\cos\frac{\pi k}{n+1}}\right).\nonumber
\end{eqnarray}
Using the formula (\ref{us1}) with $j=n$ we obtain
\[
S(a,z)=\frac{2^{-3/2}}{\left(z-\frac{a}{2}\right)^2}
\left[\frac{x_+^{2n+1}(z)-x_+(z)}
{x_+^{2(n+1)}(z)-1}-
\frac{x_+^{2n+1}(a/2)-x_+(a/2)}
{x_+^{2(n+1)}(a/2)-1}\right]
\]
\[
+\frac{2^{-1/2}}{z-\frac{a}{2}}\left[2(n+1)
\frac{x_+^{4n+2}(a/2)-x_+^{2n+2}(a/2)}
{\left(x_+^{2(n+1)}(a/2)-1\right)^2}
-\frac{(2n+1)x_+^{2n}(a/2)-1} {x_+^{2(n+1)}(a/2)-1}
\right]\frac{x_+(a/2)}{\sqrt{a^2-8}}.
\]
In particular
\[
S(3,z)=\frac{2^{-3/2}}{\left(z-\frac{3}{2}\right)^2}
\left[\frac{x_+^{2n+1}(z)-x_+(z)}
{x_+^{2(n+1)}(z)-1}- \frac{2^{n+1/2}-\sqrt{2}}
{2^{n+1}-1}\right]+
\]
\begin{equation}
+\frac{1}{z-\frac{3}{2}}
\frac{2^{2n+1}-n2^{n+1}-2^n-1}{\left(2^{n+1}-1\right)^2}.
\label{sss}
\end{equation}
Therefore for any $z>\sqrt{2}$ we obtain the following large-$n$ asymptotics
\[
\frac{N}{\Sigma_n(z)}=\frac{2\left(z-\frac{3}{2}\right)^2}{3z-4-\sqrt{z^2-2}}+O\left[x_+^{-2n}(z)+n2^{-n}
\right].
\]

If we introduce a new variable $\zeta$ via $z=\sqrt{2}(1+\zeta n^{-2})$, then we obtain the following
asymptotics
\[
\frac{N}{\Sigma_n\left(\sqrt{2}(1+\zeta n^{-2})\right)}=
\frac{\left(\sqrt{2}(1+\zeta n^{-2})-\frac{3}{2}\right)^2}
{\frac{1}{\sqrt{2}}\frac{x_+^{2n+1}(1+\zeta n^{-2})-x_+(1+\zeta
n^{-2})}{x_+^{2n+2}(1+\zeta n^{-2})-1}+\sqrt{2}(1+\zeta
n^{-2})-2}+O\left(n2^{-n}\right).
\]

\end{document}